\documentclass[prodmode,acmtecs]{acmsmall}
\usepackage{amsmath} 

\usepackage[ruled]{algorithm2e}

\usepackage{float}
\usepackage{subfigure}

\usepackage{mdwmath}
\usepackage{mdwtab}
\usepackage{algorithmic}

\newtheorem{lemma0}{\bf Lemma}

\newtheorem{definition0}{\bf Definition}

\newtheorem{theorem0}{\bf Theorem}

\renewcommand{\algorithmcfname}{ALGORITHM}
\SetAlFnt{\small}
\SetAlCapFnt{\small}
\SetAlCapNameFnt{\small}
\SetAlCapHSkip{0pt}
\IncMargin{-\parindent}

\begin{document}

\markboth{J. Sun}{Privacy-Preserving Verifiable Incentive Mechanisms
for Crowd Sensing Applications}

\title{Privacy-Preserving Verifiable Incentive Mechanisms for Crowd Sensing Applications}
\author{Jiajun Sun
\affil{Beijing University of Posts and Telecommunications, CHINA}
}

\begin{abstract}
Crowd sensing, as a new paradigm that leverages pervasive
smartphones to efficiently collect and upload sensing data, recently
has been intensively explored. Incentive mechanisms with the truthfulness 
are proposed to attract extensive users to participate so as to
achieve good service quality, enabling numerous novel applications.
Although these mechanisms are so promising, there still exist many
security and privacy challenges in real-life environments, such as
cost privacies, sensing preferences, and the payment behavior of the
platform (the crowd sensing application organizer). In this paper,
we present two privacy-preserving verifiable incentive mechanisms
for crowd sensing applications with homogeneous services,
heterogeneous services, and submodular services under the budget
constraint, not only to explore how to protect the privacy of the
users and platform, but also to ensure the verifiable correctness of
payments between the platform and users for crowd sensing
applications. Results show that our general privacy-preserving
verifiable incentive mechanisms achieve the same results as the
generic one without privacy preservation.
\end{abstract}




\acmformat{Jiajun Sun
Privacy-Preserving Verifiable Incentive Mechanism for Crowd
Sensing Applications.}

\begin{bottomstuff}
%
\end{bottomstuff}

\maketitle

\section{Introduction}\label{sec:intro}
More recently, crowd sensing have emerged as an effective and
efficient way to solve complex sensing issues \cite{mamcs2014}. For
instance, Nericell \cite{mohan2008nericell}, SignalGruru
\cite{koukoumidis2011signalguru}, and VTrack
\cite{thiagarajan2009vtrack} for providing omnipresent traffic
information, Ear-Phone \cite{rana2010ear} and NoiseTube
\cite{maisonneuve2009noisetube} for making noise maps. 
Although these crowd sensing applications have been developed,
incentive mechanisms are necessary to achieve good service quality.
Consequently, some researchers such as Singer et al, separately
propose the auction mechanisms to incentivize extensive users to
participate in crowd sensing applications so as to meet the previous
service demands
\cite{singer2010budget,singer2013pricing,yang2012crowdsourcing}.
These novel mechanisms guarantee the truthful participation of users
by determining near-optimal prices of assignments for crowd sensing
applications with the budget constraint. More importantly, the
mechanisms are incentive compatible, budget feasible, and have
competitive ratio performance and performs well in practice, thereby
ensuring these mechanisms applicable to crowd sensing applications.

Despite their merits, payments' verifiability and privacy issues,
two critical human factors in crowd sensing applications, have not
been fully explored. 
A common hypothesis made in the above mechanisms is that the
involved parties will follow the protocols honestly without the
concern of their privacy. However, some users could behave selfishly
to protect their cost privacy, sensing preferences' privacy and
identification privacy, thereby violating the hypothesis and making
these well-designed mechanisms inefficient. On the other hand, the
platform need to keep the set of current winners secrecy to maximize
his utility when facing the adversarial users. Thus, it is
imperative to provide some measures to eliminate the privacy-leakage
concerns of users and platform so as to achieve good service
quality.

In addition to the privacy issue, the payments' verifiability issue
is also a crucial human factor for the wide acceptance of the above
crowd sensing applications. It is because that some controller of
the platform (the crowd sensing organizer) may misbehave, e.g.,
provide false results or insert a fictitious bid and sensing
preference just below the payment to the users so as to deceitfully
decrease the payment to users
\cite{parkes2008practical,dong2011secure}. If the correctness of the
payments from the platform is not well guaranteed, users will be
reluctant to participate in crowd sensing applications. In practice,
since a real-world platform is operated by an individual within a
large corporation, or by a public servant within a government
department, the possibility of incorrect operations from the
platform exists in crowd sensing applications. For example, the
World Bank recently estimated the volume of incorrect exchanging
hands for public sector procurement alone to roughly US$\$200$
billion per year, with the annual volume of the procurement projects
tainted by incorrect operations close to US$\$1.5$ trillion. Thus,
how to deal with the payments' verifiability is crucial for the
success of crowd sensing applications.


Although both privacy and verifiability issues have been identified
as two crucial human factors for the wide acceptance of crowd
sensing applications, many recent research works
\cite{huang2013spring,jung2013efficient,angel2013verifiable,catanesecure}
tend to separately study them in crowd sensing applications. The
reason is that, if the privacy and verifiability issues are
addressed at the same time in crowd sensing applications, the
problem would become more challenging. For example, some privacy
enhanced techniques \cite{ganti2008poolview,shi2010prisense} enable
a user to hide his identity and sensing profile (i.e., cost and
sensing preferences like locations), but they could make some
verifiable strategies, especially the non-truthfulness incentive
strategies, hard to implement in the above truthful incentive
mechanisms, since the platform needs to greedily select winners and
compute the threshold payment based on the examination of a user's
sensing profile. However, the improvement of the verifiability needs
to reveal more information, thereby reducing privacy of users and
platform. Therefore, how to simultaneously address privacy and
verifiability issues becomes particularly challenging in crowd
sensing applications.

To tackle the above-mentioned challenges, in this paper, we present
a first step towards a crowd sensing system in which users can
verify the payments from the platform without revealing any
additional information by using the order preserving encryption
scheme (OPES) \cite{agrawal2004order}. Our approach is to enable
users to verify the payments with the help of an auction issuer
(AI): The AI chooses winners and greedily computes the threshold
payment based on encrypted user's sensing profiles. Since these
encrypted sensing profiles are order-preserving, the threshold
payment is the same as the one produced by the platform, thereby
solving the verifiability without reducing privacy of users and
platform. Specifically speaking, we first introduce three incentive
mechanisms for crowd sensing applications with homogeneous sensing
jobs, heterogeneous sensing jobs and submodular sensing jobs (to be
elaborated later). Then, we propose a general privacy preservation
verifiable incentive mechanism for homogeneous sensing jobs and
heterogeneous sensing jobs. Furthermore, we also propose a privacy
preservation verifiable incentive mechanism for submodular sensing
jobs. The two mechanisms are implemented by introducing the
oblivious transfer (OT), the timed lapse cryptography services
(TLC), and the bulletin board, satisfying the three desirable
properties: the non-repudiation by users and the platform, secrecy,
and verifiable correctness. Finally, analysis show that our
privacy-preserving verifiable incentive mechanisms achieve the same
results as the generic one without privacy preservation and
verification.

The rest of the paper is organized as follows. In Section
\ref{related}, we briefly discuss the related work and motivation.
In Section \ref{SystemModel}, we present our relative models and our
design goal. In Section \ref{MecahnismDesgn}, we introduce novel
incentive mechanisms for crowd sensing applications with the budget
constraint. Based on these mechanisms, in Section
\ref{privacymechanisms}, we design two privacy-preserving verifiable
incentive mechanisms satisfying the above three desirable
properties, followed by the security analysis and performance
evaluation in Section \ref{Analysis} and Section \ref{Experiment}.
Finally, we draw our conclusions in Section \ref{Conclude}.
\section{Background and Related Work}~\label{related}

Privacy-preserving mechanisms have been extensively explored in
crowd sensing applications. Most of these research works are based
on $k$-anonymity \cite{sweeney2002k}, where a user's location is
cloaked among $k-1$ other users. For instance, the authors of
\cite{kalnis2007preventing} and \cite{gedik2008protecting} use the
temporal and spatial cloaking techniques to preserve users' privacy.
Their mechanisms blind the participant's location at a specific time
in a cloaked area to achieve the privacy requirements. The authors
of \cite{shilton2008participatory,shin2011anonysense,de2011short}
study the privacy protection in crowd sensing applications by
applying a privacy regulation technique. Furthermore, the authors of
\cite{shin2011anonysense} and \cite{de2011short} focus on how users
submit the jobs to the platform without disclosing their identity.
Different from the above anonymous collection mechanisms, the
authors of \cite{huang2013spring} protect the privacy of users by
applying the OT \cite{rabin1981exchange}. However, they do not
consider the verifiability of user's inputs and outcomes.

Additionally, verifiability of the payment is also a vital factor an
incentive mechanism design faced. The verifiability of payments have
been extensively explored in traditional auction mechanisms. For
instance, the authors of \cite{naor1999privacy,juels2003two} apply
the proxy OT to verify the payment of the platform by constructing a
circuit. The authors of \cite{parkes2008practical} use a timed lapse
cryptography service to keep users' bids secret from the platform
before the auction closed, and prevent them from rigging their bids
after bidding. However, they do not apply for crowd sensing
applications, since they neglect the effect of a large of
participants in crowd sensing applications. Recently, a timed
commitment encryption method is adopted to enhance the level of the
payment correctness from the platform for crowd sensing
applications. For example, the authors of
\cite{catanesecure,zhao2012private,angel2013verifiable} apply the
timed commitment to tackle the verifiable correctness issue in
different aspects. However, these mechanisms are not applicable in
real crowd sensing applications with the budget constraint.

In this paper, to solve the above challenges, we introduce the
bulletin board, OT, and TLC to guarantee the privacy and
verifiability for crowd sensing applications without sacrificing the
platform's utility and truthfulness.

\section{System Model and Problem Formulation}~\label{SystemModel}

In this section, we first expound our system model, auction model,
adversarial model, and the bulletin board applied to our
privacy-preserving verifiable incentive mechanisms. Then we present
our goal for crowd sensing applications.

\subsection{System Model}
\begin{figure}
\setlength{\abovecaptionskip}{0pt}
\setlength{\belowcaptionskip}{10pt} \centering
\centering
\includegraphics[width=0.52\textwidth]{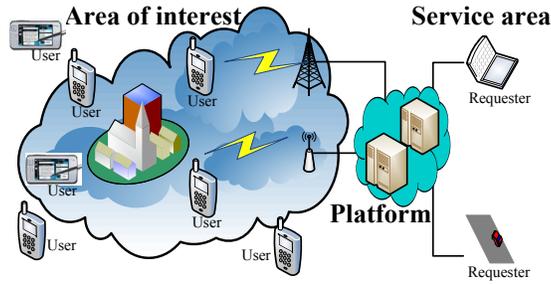}
\caption{Our crowd sensing system model.} \label{crowd}
\end{figure}

We consider the following system model for crowd sensing
applications, illustrated in Fig.~\ref{crowd}. The system consists
of a crowd sensing platform that resides in the cloud, a requester,
and many mobile device users that are connected to the platform by
cellular networks (e.g., GSM/3G/4G) or WiFi connections. The
requester posts a crowd sensing task with a budget $B>0$ to the
platform. There are $m$ available assignments in each task.
Receiving the task, the platform publicizes a crowd sensing campaign
towards the area of interest (AoI), aiming at finding some users to
maximize the number of assignments performed efficiently. Assuming
that
a set of users $\mathcal{U}=\{1,2,\cdots,n\}$ in the AoI is interested in the campaign. 
In this paper, with respect to the model of sensing jobs completed
by all users, we discuss the following three sensing job models
proposed in \cite{singer2010budget} for the crowd sensing campaign:

\textbf{Homogeneous sensing job model:} Both each sensing job
assignment and the limit of the number of assignments completed by
each user are the same. Meanwhile, each user can complete only a
single assignment.

\textbf{Heterogeneous sensing job model:} Each sensing job
assignment is the same, but the limit of the number of assignments
completed by each user is different. It means that different users
can complete different number of assignments.

\textbf{Submodular sensing job model:} Each sensing job assignment
is different, and each user $i$ can do a subset $\Gamma_{i}$ of
assignments $\Gamma$.

If the campaign is oriented to users with the homogeneous sensing
job model and the heterogeneous sensing job model, receiving the
campaign, each user $i$ synchronously submits his sensing profile
$\mathcal{P}_{i}=(b_{i},l_{i})$, where $b_{i}$ is obtained based on
a true cost $c_{i}$ for performing a single assignment and $l_{i}$
is a limit for the number of assignments he is willing to work on.
This means that if he is a winner, at most $l_{i}$ assignments can
be allocated to him and the payment for each assignment must exceed
$b_{i}$. In this case, the sensing job model is the heterogeneous
sensing job model, which indicates that different users can complete
different number of assignments. When $l_{i}=1$, the sensing job
model is reduced to the homogeneous sensing job model, which
indicates that each user can complete only a single assignment.

If the campaign is oriented to users with the submodular sensing job
model, receiving the campaign, each user $i$ synchronously submits
his sensing profile $\mathcal{P}_{i}=(b_{i},\Gamma_{i})$, where
$b_{i}$ is obtained based on a true cost $c_{i}$ for performing the
sensing service with his assignments' set $\Gamma_{i}$, i.e.,
$\Gamma_{i}\subseteq\Gamma=\{\tau_{1},\tau_{2},\cdots,\tau_{m}\}$.
We assume that $l_{i}$ or $\Gamma_{i}$ is fixed. Furthermore, 
%
under the budget constraint $B$, the platform, when presented with
the sensing profiles of all users, must decide a subset of users to
select, and how much payment to pay to each selected user. Our goal
is to make incentive mechanisms to achieve non-repudiation by users
and platform, secrecy, and verification without sacrificing the
above standard economic goal such as utility maximization,
truthfulness.

\subsection{Auction Model}
\begin{figure}
\setlength{\abovecaptionskip}{0pt}
\setlength{\belowcaptionskip}{10pt} \centering
\centering
\includegraphics[width=0.43\textwidth]{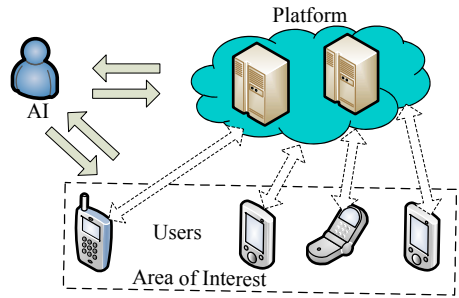}
\caption{Our general crowd sensing auction scenario.}\label{auction}
\end{figure}

We model the above interactive process as a sealed-bid auction
between the platform and users (see Fig.~\ref{auction}), in which
there is a crowd sensing platform, a set of participatory users, and
an AI. A set of assignments $\Gamma$ publicized by the platform
during the deadline $T$ are auctioned towards $n$ users in crowd
sensing applications. Each user $i$ submits his encrypted sensing
profile $\mathcal{P}_{i}$, i.e., a pair of encrypted $b_{i}$ and
encrypted $l_{i}$ or $\Gamma_{i}$. The AI is semi-honest (passive or
curious), and only checks the platform randomly. This will be
further illustrated in Section \ref{MecahnismDesgn}.

\subsection{Bulletin Board}
The platform maintains a bulletin board. It can be a publicly known
website maintained and updated by the platform. The platform applies
the bulletin board to post all public information about the
mechanism, including all abstractions (e.g., some auction details)
as well as encrypted information about users' profiles, the methods
of the winner selection and payment determination that can be used
to verify the payment correctness from the platform. All posted to
the bulletin board will carry appropriate digital signatures so as
to identify their originators. For the ease of exposition, we will
refer to $sign_{i}(m)$ as the signature of the message $m$ from the
user $i$ in the rest of this paper. All abstractions can be
constructed by using standard cryptographic techniques. We calculate
the communication overheads of our mechanisms as the number of
auction details published on the bulletin board. It is worth noting
that a robust bulletin board is needed for crowd sensing
applications. Thus our mechanisms can just exploit standard
broadcast techniques.

\subsection{Adversarial Model}
In the auction process with the budget constraint, the platform is
supposed to know only the set of current winners, and their sensing
profiles. Each user $i$ only learns whether he is the winner, and he
is paid if he is a winner. He does not know anything about others'
profiles except for the very limited implicit information in the
payment from the platform.

We assume that the platform and users are semi-honest adversaries in
our mechanisms, and collusion of bidders and platform does not
exist. That is, the platform is interested in inferring each user's
private information no matter he is a winner or not. Users try to
infer others' sensing profiles to maximize their own utilities.
Besides, the platform and users can also collude with each other.
According to the above auction model, we give the analysis of the
privacy in our framework below.

\begin{definition0}\label{df:wardropequilibrium}
Given all the communication strings $\mathcal{C}$ and its output of
the auction $Output$ during the auction, an adversary's advantage
over the privacy information $\zeta_{i}$ of user $i$ is defined as
$Adv_{\zeta_{i}}=P_{r}[\zeta_{i}|\mathcal{C},
Output]-P_{r}[\zeta_{i}|Output]$, where $P_{r}[\zeta_{i}]$ is the
probability that a correct $\zeta_{i}$ is inferred. In this paper,
$\zeta_{i}$ can be the bid or sensing services $\Gamma_{i}$ of user
$i$.
\end{definition0}

\begin{definition0}\label{df:privacymodelkanonymity}
A privacy-preserving scheme satisfies $k$-anonymity, if a user
cannot be identified by the sensitive information with probability
higher than $1/k$ \cite{sweeney2002k}.
\end{definition0}

In this paper, our security goal is to achieve a scheme such that
the advantage is of a negligible function of the security parameter
or $k$-anonymity is guaranteed.

\subsection{Problem Formulation}
According to the above sensing job model, we need to consider two
cases. One case is when sensing job model is the homogeneous sensing
job model or the heterogeneous sensing job model. In the two models,
the platform needs to design a mechanism $\mathcal{M}=(f,p)$, which
consists of an allocation function $f:
\mathcal{R}_{+}^{n}\rightarrow \mathcal{Z}_{+}^{[n]}$ and a payment
function $p: \mathcal{R}_{+}^{n}\rightarrow\mathcal{R}_{+}^{n}$. The
allocation function $f$ maps a set of $n$ bids to an allocation of
assignments for a selected subset of users. In particular, in the
homogeneous sensing job model, if user $i$ is selected, $f_{i}=1$.
The payment function $p$ returns a vector $(p_{1},\cdots, p_{n})$ of
payments to the users. That is, each selected user $i\in S$ is
allocated $f_{i}$ assignments at price $p_{i}$ per task. In the
heterogeneous user model, the utility of user $i$ is
$f_{i}(p_{i}-c_{i})$ if it is selected, i.e., $i\in S$, $0$
otherwise. The existing goal of the platform is to maximize the
number of assignments under given budget $B$, i.e., $\max\sum_{i\in
S} f_{i}$, subject to $\sum_{i\in S}f_{i}p_{i}\leq B$, $\forall i,
f_{i}\leq l_{i}$. In particular, when $l_{i}=1$, the above results
are also applicable to the homogeneous sensing job model.

The other case is when sensing job model is the submodular sensing
job model. The platform needs to design a mechanism
$\mathcal{M}=(f,p)$, which consists of an allocation function $f:
\mathcal{R}_{+}^{n}\rightarrow 2^{[n]}$ and a payment function $p:
\mathcal{R}_{+}^{n}\rightarrow\mathcal{R}_{+}^{n}$. The allocation
function $f$ is a indicator function that returns $1$ if user $i$ is
allocated and $0$ otherwise. The utility of user $i$ is
$p_{i}-c_{i}$ if it is selected, i.e., $i\in S$, $0$ otherwise. The
payment function $p$ returns a vector $(p_{1},\cdots, p_{n})$ of
payments to the users. The existing goal of the platform is to
maximize the value from the selected users' services under the
budget constraint $B$, i.e., $\max V(S)$, subject to $\sum_{i\in
S}p_{i}\leq B$, where $V(S)$ is monotone submodular.

However, the above maximal problem will bring many security and
privacy issues including users' sensing profiles. Since users are
reluctant to disclose all these private information to others as
well as the platform. On the other hand, both the winners and the
platform should be able to verify the payment provided by our
mechanisms. Thus, beyond the standard economic goals (e.g.,
truthfulness, individual rationality, utility maximization, etc.),
our mechanisms also satisfy the following three desirable
properties:
\begin{itemize}
   \item \textbf{Non-repudiation by users and platform:} For each
   user, once it submits a encrypted profile, its encrypted profile is provably
   unalterable. For the platform, its exclusion of a properly
   submitted encrypted profile can be conclusively proven. Thus, it must
   action legally.
   \item \textbf{Secrecy:} These encrypted profiles are bidden to all
   users and the platform until their holders become winners of
   mechanisms. When they become winners, only the platform knows
   their decrypted profiles. When they become losers, their encrypted profiles are bidden to
   all other
   users and the platform. The current winners' profiles of the platform are bidden to all users no matter whether these
   users are winners or not.
   \item \textbf{Verifiable correctness:} Since the mechanism itself is truthful, we only need
to guarantee that each user can verify the payment correct provided
from the platform by applying the computation method published on
the bulletin board.
\end{itemize}

\section{Incentive Mechanisms for Crowd Sensing Applications} \label{MecahnismDesgn}
In this section, we introduce three incentive mechanisms for crowd
sensing applications with homogeneous sensing jobs, heterogeneous
sensing jobs, submodular sensing jobs respectively. In essence, the
incentive mechanisms for crowd sensing applications require the
truthfulness, computationally effectiveness, budget feasibility and
approximation. Singer et al. present these mechanisms meeting the
four conditions well. For the ease of presentation, in the following
section, we propose two privacy preservation verifiable incentive
mechanisms based on the three mechanisms, but our mechanisms are
easy to be extended to other truthful incentive mechanisms for real
crowd sensing environments. Furthermore, their truthful mechanisms
are illustrated as follows.

To better understand the following three incentive mechanisms, let
us see the following familiar example. Given a budget constraint $B$
and subsets $\mathcal{U}=\{1,2,\cdots,n\}$ of some ground set, where
each user $i$ corresponds to a subset of the ground set and a
associated cost $c_{i}$ find a users' subset $S$ which maximizes
$|\cup_{i\in S} \{i\}|$ under the budget constraint. This is a
typical coverage problem, called submodular sensing job model here,
in which each user's value depends on the identity of the sensing
data set it holds. When each user's value only depends on the
cardinality of the sensing data set, rather than the identity of the
sensing data set it holds, it means that different users can
complete different number of sensing data, thereby simplifying the
submodular sensing job model to heterogeneous sensing job model.
Furthermore, if each user only completes a single sensing data
assignment, the heterogeneous sensing job model will become a
homogeneous sensing job model. For the simplicity of presentation,
we first introduce the incentive mechanism with homogeneous sensing
job model.

\subsection{Incentive Mechanism with Homogeneous Sensing Jobs}
For crowd sensing applications with
homogeneous sensing jobs, consider the above-mentioned allocation
rule $f$: Sort the $n$ bids reported by $n$ users so that $b_{1}\leq
b_{2}\leq \cdots\leq b_{n}$, and find the largest $k$ so that
$b_{k}\leq B/k$. That is, $k$ is the place where the curve of the
increasing costs intersects the hyperbola $B/k$.The set allocated
here is $\{1,2,\cdots,k\}$. That is, winners' set
$S=\{1,2,\cdots,k\}$. This is obviously a monotone allocation rule:
a user can be not excluded when decreasing his bid. In
\cite{singer2010budget}, the authors design the following incentive
mechanism for crowd sensing applications with homogeneous sensing
jobs and show the mechanism satisfies the above four conditions.


More formally, firstly, sorting the users' bids: satisfying
$b_{1}\leq b_{2}\leq \cdots\leq b_{n}$. Then find the largest
integer $k$ such that $b_{k}\leq B/k$. Finally, determine the set of
allocated users to be $S=\{1,2,\cdots,k\}$, and provide same payment
$p_{i}=\min\{B/k, b_{k+1}\}$.

\subsection{Incentive Mechanism with Heterogeneous Sensing Jobs}
For crowd sensing applications with heterogeneous sensing jobs, the
authors of \cite{singer2013pricing} present the following mechanism
for determining near-optimal prices of jobs for crowd sensing
applications with heterogeneous sensing jobs. Their mechanism is
illustrated as follows: Firstly, sort the users' bids: satisfying
$b_{1}\leq b_{2}\leq \cdots\leq b_{n}$. Then find the largest
integer $k$ such that $b_{i}\leq B/{\sum\nolimits_{j\leq i}
{f_{j}}}$. Finally, determine the set of allocated users
$S=\{1,2,\cdots,k\}$, and provide the same payment
$\min\{B/\sum\nolimits_{j \leq i} {f_j }, b_{i+1}/l_{i+1}\}$ for
completing a sensing job.

Obviously, the homogeneous user model is a special case of the
heterogeneous user model, i.e., $l_{i}=1$ for each user $i$.
\subsection{Incentive Mechanism with Submodular Sensing Jobs} For
crowd sensing applications with submodular sensing jobs, the authors
of \cite{singer2010budget,tran2012efficient,yang2012crowdsourcing}
apply the proportional share allocation rule proposed in
\cite{singer2010budget} to address the extensive user participation
issue for crowd sensing applications, which consists of two phases:
the winner selection phase and the payment determination phase. We
first introduce definition of the submodular utility function.

\begin{algorithm}[htb] 
\renewcommand{\algorithmicrequire}{\textbf{Input:}}
\renewcommand\algorithmicensure {\textbf{Output:} }
\caption{An auction mechanism for submodular sensing jobs under the budget constraint} 
\label{originalbudgetauction} 
\begin{algorithmic}[1] 
\REQUIRE User set $\mathcal{U}$, the budget
constraint $\mathcal{B}$. \\
\ENSURE The set of winners  $S$. \\
\STATE // Phase 1: Winner selection \\
\STATE $S\leftarrow \emptyset$; $i\leftarrow\arg\max_{j\in\mathcal
{U}}U_{j}(S)/b_{j}$;\\

\WHILE{$U_{i}(S)/b_{i}\geq U(S\cup i)/B$}
                    \STATE $S\leftarrow S\cup i$;\\
                    \STATE $i\leftarrow\arg\max_{j\in\mathcal {U}\setminus S}(U_{j}(S)/b_{j})$;\\
            \ENDWHILE
\STATE // Phase 2: Payment determination \\
\FOR{each user $i\in \mathcal{U}$} \STATE $p_{i}\leftarrow 0$;\\
\ENDFOR
\FOR{each user $i\in S$} \STATE $\mathcal{U}^{'}\leftarrow \mathcal{U}\backslash\{i\}$; $\mathcal{T}\leftarrow \emptyset$;\\
\REPEAT
   \STATE $i_{j}\leftarrow \arg \max_{j\in \mathcal{U}^{'}\backslash \mathcal{T}}(U_{j}(\mathcal{T})/b_{j})$;\\
   \STATE $p_{i}\leftarrow \max \{p_{i}, \min \{b_{i(j)}, \eta_{i(j)}\}\}$;\\
   \STATE $\mathcal{T}_{j-1}\leftarrow \mathcal{T}$;
   $\mathcal{T}\leftarrow \mathcal{T}\cup \{i_{j}\}$;\\
\UNTIL {$b_{i_{j}}> U_{i(j)}(\mathcal{T}_{j-1})B/U(\mathcal{T})$}
    \ENDFOR
 \RETURN ($S$, $p$);\\

\end{algorithmic}
\end{algorithm}

\begin{definition0}[\textbf{Submodular Function}]\label{df:submodule}
Let $\mathbb{N}$ be a finite set, a function $U$ :
$2^{\Omega}\rightarrow \mathbb{R}$ is submodular if
$U(S\cup\{i\})-U(S)\geq U(T\cup\{i\})-U(T), \forall S\subseteq
T\subseteq \Omega$, where $\mathbb{R}$ is the set of reals.
\end{definition0}

From the above Definition \ref{df:submodule}, we can know the
utility function $U$ is submodular and derive the following sorting
according to increasing marginal contributions relative to their
bids from users' set to find the largest $k$ satisfying $b_{k}\leq
U_{k}B/U(S\cup k)$.

\begin{equation}\label{marginaleq}
U_{1}/ b_{1}\geq U_{2}/ b_{2}\geq \cdots \geq U_{|\mathcal{U}|}/
b_{|\mathcal{U}|},
\end{equation}
where $U_{k}$ denotes $U_{k\mid S_{k-1}}$
($=U(S_{k-1}\cup\{k\})-U(S_{k-1})$), $S_{k}=\{1,2, \cdots,k\}$, and
$S_{0}=\emptyset$. 
To calculate the payment of each user, we sort the users in
$\mathcal{U}\backslash\{i\}$ similarly as follows:

\begin{equation}\label{marginaleq}
U_{i_{1}}(\mathcal{T}_{0})/ b_{i_{1}}\geq
U_{i_{2}}(\mathcal{T}_{1})/ b_{i_{2}}\geq \cdots \geq
U_{i_{n-1}}(\mathcal{T}_{n-2})/ b_{i_{n-1}},
\end{equation}

The marginal value of user $i$ at the position $j$ is
$BU_{i(j)}(\mathcal{T}_{j-1})/U(\mathcal{T}_{j})$. Assume that
$k^{'}$ to be the position of the last user
$i_{j}\in\mathcal{U}\backslash\{i\}$, such that $b_{i_{j}}\leq
U_{i(j)}(\mathcal{T}_{j-1})B/U(\mathcal{T})$. To guarantee the
truthfulness, each winner should be given the payment of the
critical value. This indicates that user $i$ can not win the auction
if it reports higher than this critical value. More details are
given in Algorithm \ref{originalbudgetauction}, where
$b_{i(j)}=U_{i(j)}(\mathcal{T}_{j-1})b_{i_{j}}/U_{i_{j}}(\mathcal{T}_{j-1})$
and
$\eta_{i(j)}=U_{i(j)}(\mathcal{T}_{j-1})B/U(\mathcal{T}_{j-1}\cup\{i\})$.

However, although the above three mechanisms under the given budget
constraint are so promising for crowd sensing applications, it also
bring many verifiability and privacy issues including users' sensing
profiles and the payment correctness. 
In the following section, we will design two privacy-preserving
verifiable incentive mechanisms for crowd sensing applications with
homogeneous and heterogeneous sensing jobs, and submodular sensing
jobs to address the above-mentioned challenges.

\section{Design Details}~\label{privacymechanisms}

During the above three auction mechanisms for crowd sensing
applications with homogeneous, heterogeneous jobs and submodular
jobs, we need to choose a set of winners and finish the payment of
the winners according to the mechanism with the given budget
constraint. In this section, we first introduce basic cryptographic
schemes. Then we apply the schemes to design our privacy
preservation verifiable auction mechanism for homogeneous,
heterogeneous jobs and submodular jobs respectively.

\subsection{Basic Cryptographic Schemes}\label{basictools}

In this section, we introduce the constructions of time-lapse
cryptography service and blinded digital signature for achieving the
goal of non-repudiation by users and platform, OT for making users'
sensing profile secret, and the computation of marginal utility and
set union for making platform' current winners' set secret. Then, in
the following details, we apply the bulletin board and the parameter
$\alpha$ to ensure verifiability of payments and the payment
correctness.

\subsubsection{Time-Lapse Cryptography Service}
In our proposed mechanism we apply timed commitments on sensing
profiles of all users until the auction closes. Cryptographic
methods, as presented in \cite{boneh2000timed} can be used to
implement the timed-commitments. Considering the computation
efficiency reasons, we choose a time lapse cryptography (TLC)
service from \cite{rabin2006time}, which makes it possible to use
commitments with the classical hiding and binding properties.
Besides, it prevents users from refusing to reveal committed sensing
profiles and also preventing the platform from dropping received
commitments, claiming not to have been able to reveal the committed
sensing profiles. In our mechanisms, an auction issuer (AI), acting
as the TLC service provider, publishes a public key of a
non-malleable encryption scheme, and sends the corresponding private
key only when the auction closes. Whenever timed commitments on
sensing profiles are applied, it means that a user encrypts her
sensing profile by applying the AI-generated public encryption key.
Besides, receiving the corresponding private key, the platform can
know the encrypted sensing profile.

\subsubsection{Blinded Digital Signature}
In our work, each user is a signer who is introduced only to keep
the confidentiality of its the following transformed bid and sensing
subset of assignments to the platform as well as other users.
Considering the security, not all digital signature schemes can be
used. To these goals, we apply the Nyberg-Rueppel signature scheme
\cite{camenisch1995blind} (see Algorithm \ref{NebergSignature}).
Notably, we do not need the signer to verify the authenticity of
them, and on the other hand the platform can obtain their
transformed bids and sensing preference selections from all signers.
For the ease of exposition, we will refer to $sign_{i}(m)$ as the
signature of the message $m$ from the user $i$ in the rest of this
paper. Note that the signature scheme requires the message to be an
integer, therefore, we need to apply $sign(\lfloor10^{k}m\rfloor)$
for the input $m$ if $m$ is not an integer like the bid, where $k$
can be appropriately chosen to preserve the rank from
$\{3,4,\cdots\}$ and $\psi(x)$ denotes the output of the signature
scheme. At the same time, we remove the signature by using
$10^{-k}sign^{-1}(c_{m})$, where $c_{m}$ is obtained by the
signature $c_{m}=sign_{i}(m)$. For ease of exposition, in the rest
of the paper, we assume that the value of the signature is an
integer. According to \cite{jung2013efficient}, the deviation for
the roundness of the signature is negligible. Thus, our assumption
is reasonable.
\begin{algorithm}[h] 
\renewcommand{\algorithmicrequire}{\textbf{Input:}}
\renewcommand\algorithmicensure {\textbf{Output:} }
\caption{Blinded Nyberg-Rueppel Signature.} 
\label{NebergSignature} 
\begin{algorithmic}[1] 
\STATE Initialize a prime number $p$, a prime factor $q$ of $p-1$,
and an element $g \in {\rm \mathbb{Z}}_p^*$ with order $q$;\\
\STATE The signer selects $\tilde{k}\in {\rm \mathbb{Z}}_p$ and
sends $\tilde{r}=g^{\tilde{k}}$(mod $p$) to signee;\\
\STATE The signee randomly chooses $\alpha\in {\rm \mathbb{Z}}_q$,
$\beta \in {\rm \mathbb{Z}}_q^*$, computes $r=mg^{\alpha}$(mod $p$)
and $\tilde{m}=r\beta^{-1}$(mod $q$) until $\tilde{m}\in {\rm
\mathbb{Z}}_q^*$. Then, he sends $\tilde{m}$ to the signer;\\
\STATE The signer computes $\tilde{s}=\tilde{m}x +\tilde{k}$(mod
$q$) and sends $\tilde{s}$ to the signee; \\
\STATE The signee computes $\tilde{s}=\tilde{s}\beta +\alpha$(mod
$q$), and the pair ($r,s$) is the the signature for $m$; \\
\STATE Check whether $m=g^{-s}y^{r}r$(mod $q$) to verify the correctness.\\
\end{algorithmic}
\end{algorithm}

\subsubsection{OT for Privacy Preservation}\label{OTtech}
OT is a paradigm of secret exchange between two parties, users and a
platform. Each user can achieve one of $n$ secrets from the user,
without knowing any information about the rest of $n$ secrets, while
the platform has no idea which of the $n$ secrets is accessed. Our
work employs an efficient $1$-out-of-$z$ OT of integers
\cite{tzeng2004efficient}. The detailed description is given in the
Algorithm \ref{OT}.

\begin{algorithm}[h] 
\renewcommand{\algorithmicrequire}{\textbf{Input:}}
\renewcommand\algorithmicensure {\textbf{Output:} }
\caption{Oblivious Transfer ($OT_n^1$).} 
\label{OT} 
\begin{algorithmic}[1] 
\STATE Initialization: System parameters: $(g,h,G_{g})$; the AI's
input: $m _1 ,m _2 , \cdots ,m _n\in G_{g}$; user
$i$'s choice: $\alpha, 1\leq\alpha\leq n$;\\
\STATE User $i$ sends $y=g^r h^\alpha$;\\
\STATE The AI replies with $c_{i}=(g^{k_i } ,m_i (y/h^i )^{k_i })$, $k_i\in_{R}Z_{q}$, $1\leq i\leq n$.\\
\STATE By $c_{\alpha}=(a,b)$, user $i$ computes $m_\alpha=b/a^r$;\\
\end{algorithmic}
\end{algorithm}
\subsubsection{Marginal Utility Computation}\label{marginalcomputation}
Besides the above losers' sensing preferences, the current winner
set $S$ produced by the platform, should be also kept secret to all
users. In such problems, how to compute the marginal utility without
knowing $S$ is challenging. We address it by introducing
multivariate polynomial evaluation protocol
(MPEP)\cite{jung2013privacy,zhang2013verifiable}, in which the
multivariate polynomial are computed without disclosing any $x_{i}$
input of various users as follows: $f(\vec x) = \sum\nolimits_{k =
1}^m {(c_k \prod\limits_{i = 1}^n {x_i^{d_{i,k} } } )}$, where there
is a group of open $m$ powers for each user and $m$ coefficients to
any participant as well as the attackers. We compute the marginal
utility by assuming that there are $m$ sensing data points and $n$
mobile users. Then we have $m$-dimensional vector $C_{S}$ indicating
whether $m$ sensing data points are included in currently chosen
sets $S$, where $c_{k,S}=1 - \prod\nolimits_{j = 1}^n {(1 -
c_{j,k,S})}$. If $k$-th data point is in user $j$'s subset
$\Gamma_{j}$ of assignments and user $j$ is in $S$, $c_{k,S}=1$, and
0 otherwise. Since each user knows whether it belongs to $S$, each
winning user's marginal utility can be evaluated via one aggregator
MPEP with the help of $n$ users and only user $i$ receives the
result by applying the above MPEP equation. Finally, the user $i$
can divide his bid $b_{i}$ to the result to compute the
marginal-utility-per-bid value $\omega_{i}$. The detailed expression
is given as follows:
\begin{equation*}
\begin{split}
\omega _i &=\frac{1}{b_i} ({\sum\nolimits_{j = 1}^m {c_{j,S \cup \{
i\} } }  - \sum\nolimits_{j = 1}^m {c_{j,S} } })\\
&= \frac{1}{b_i}({\sum\nolimits_{k = 1}^m {(1 - \prod\nolimits_{j =
1}^n {(1 - c_{j,k,S \cup \{ i\} } } )} ) - \sum\nolimits_{k = 1}^m
{(1 - \prod\nolimits_{j = 1}^n {(1 - c_{j,k,S} } )} )}).
 \end{split}
 \end{equation*}

\subsubsection{Privacy Preservation Set Union Computation for Platform}
Since the current winner set $S$ is required to be kept secret to
all users, for the platform, how to compute the set union without
leakage the its privacy, i.e., the current winner set $S$ is a
challenging issue. In the paper, we address it by using Paillier
cryptosystem \cite{paillier1999public} to the set union computation.
About the set union computation, we refer interested readers to the
paper \cite{frikken2007privacy}. The detailed description is
illustrated in Algorithm \ref{setunion}. The Paillier cryptosystem
as well as its homomorphic property is also shown below:

\begin{equation}\label{homomorphicproperty}
\begin{array}{l}
 E(m_1 ,r_1 ) \cdot E(m_2 ,r_2 ) = E(m_1  + m_2 ,r_1  + r_2 ) \\
 E(m_1 ) \cdot g^{m_2 }  = E(m_1  + m_2 ,r_1 ) \\
 E(m_1 ,r_1 )^{m_2 }  = E(m_1  \cdot m_2 ,r_1  \cdot m_2 ) \\
 \end{array}
\end{equation}

\begin{algorithm} 
\renewcommand{\algorithmicrequire}{\textbf{Input:}}
\renewcommand\algorithmicensure {\textbf{Output:} }
\caption{Privacy-preserving set union computation.} 
\label{setunion} 
\begin{algorithmic}[1] 
\STATE Initialize system parameter: two same-length prime numbers $p$,$q$, public keys $n=pq$, $g \in \mathbb{Z}^{*}_{n^2}$, private key $\lambda=(p-1)(q-1)$, $\mu=\lambda^{-1}$mod $n$;\\
\STATE The platform computes the polynomial $f_{A}$ and sends the
encrypted $E_{p}(f_{S})$ to the user $u_{i}$;\\
\STATE Upon receiving $E_{p}(f_{S})$, the user $u_{i}$ chooses a
random value $r$ (choose uniformly) and computes a tuple
$(E_{p}(f_{S}(\tau)\ast \tau\ast r),E_{p}(f_{S}(\tau)\ast r))$ for
each assignment value $\tau \in S$. He randomly permutes all of the
tuples and sends them to the platform;\\
\STATE For each tuple $(E_{p}(x),E_{p}(y))$, the platform decrypts
$x$ and $y$. If both values are $0$, then the platform continues to
next tuple. Otherwise, the platform finds a good with the value
$x\ast y^{-1}$ and adds it to the output set; As such, the marginal
utility of the user $u_{i}$ can be obtained. \\
\end{algorithmic}
\end{algorithm}


\subsection{Design Privacy-Preserving Details for Homogeneous and Heterogeneous jobs}

\subsubsection{Initialization}
The platform invites the AI to participate in the auction and sends
the following information to him: the crowd sensing task identifier
$TID$ of the platform, the deadline $T$, and the timed-lapse
encryption key $TPK$ to be used by all users in commitments. If the
AI accepts them, he sets the probability of the auditions from users
as $\alpha$ so that $\alpha\geq p_{max}/(F+p_{max})$, where
$p_{max}$ and $F$ are the maximal payment and fine paid from the
platform respectively, and sends signed $\alpha$ and signed auction
details to the platform. If the platform accepts it, the platform
posts them on the bulletin board. Finally, our mechanism also
defines a set of possible bids as $\beta = \left\{ {\beta _1 ,\beta
_2 , \cdots ,\beta _z } \right\}$ and a set of possible limits of
the number of assignments, $\chi = \left\{ {\chi _1 ,\chi _2 ,
\cdots ,\chi _v } \right\}$, where $\beta _1 < \beta _2 < \cdots <
\beta _z$ and $\chi _1 < \chi _2 < \cdots < \chi _v$ hold, and
requires each user $i$'s bid $b_{i}\in \beta$ and the limit of the
number of assignments $l_{i}\in \chi$. The AI maps each bid value
$\beta _i$ and limit value $\chi _i$ to $\gamma _i$ and $\tau _i$
respectively, while preserving the rank, i.e., satisfying $\gamma _1
< \gamma _2 < \cdots  < \gamma _n$ and $\tau _1 < \tau _2 < \cdots <
\tau _n$. Similarly, users' bids and limits are transformed by using
the OPES for preserving their ranks. To this end, we assume that the
above AI can bootstrap the crowd sensing market application. All of
the above data are posted on the bulletin board, accompanied by the
platform's signature $sign_{p}$.

\begin{algorithm}[t]
\renewcommand{\algorithmicrequire}{\textbf{Input:}}
\renewcommand\algorithmicensure {\textbf{Output:} }
\caption{PVI-H// Privacy-preserving Verifiable incentive mechanism
for Crowd sensing applications with homogeneous sensing jobs or
Heterogeneous sensing jobs} 
\label{winnerpaymentselection} 
\begin{algorithmic}[1] 
\REQUIRE User set $\mathcal {U}$, the budget
constraint $B$. \\
\ENSURE $S$. \\
// Phase 1: Winner selection \\
\STATE Initialize: Each user $i$ receives his encrypted sensing
profile ($\tilde{b}_{i}$, $\tilde{t}_{i}$) by using Algorithm
\ref{OT} and submits their commitments to the platform; At time
$T+1$, the platform makes a decommitment and sorts users in
$\mathcal {U}$ i.e.,
$\tilde{b}_{1}<\tilde{b}_{2}\cdots<\tilde{b}_{|\mathcal{U}|}$;$S\leftarrow \emptyset$; $i=1$;\\

\STATE $b_{1}\leftarrow OPENS^{-1}(\tilde{b}_{1})$;\\

\WHILE{$b_{i}\leq B/{\sum\nolimits_{j\in S} {f_{j}}}$}

                    \STATE $f_{i}\leftarrow 1$;\\
                    \IF {jobs are heterogeneous}  
                        \STATE $f_{i}\leftarrow\min\{OPENS^{-1}(\tilde{l}_{i}), \tau_{i}\}$, where $\tau_{i}=\lfloor(B-b_{i}\sum\nolimits_{j \in S}
{f_j })/b_{i}\rfloor$;\\
                    \ENDIF
                    \STATE $S\leftarrow S\cup i$;\\
                    \STATE $i\leftarrow i+1$;\\
                    \STATE $b_{i}\leftarrow OPENS^{-1}(\tilde{b}_{i})$;\\
            \ENDWHILE\\
// Phase 2: Payment determination \\
\FOR{each user $i\in S$}
    \STATE $l_{i+1}\leftarrow OPENS^{-1}(\tilde{l}_{i+1})$;\\
     \IF {$j\leq i-1$}
        \STATE Pay $p_{j}f_{j}$ to user $j$;\\
    \ENDIF
    \IF {$j==i$}
        \STATE $p_{j}\leftarrow \min\{B/\sum\nolimits_{j\in S}{f_j }, b_{i+1}/l_{i+1}\}$; Pay $p_{j}f_{j}$ to user $j$;\\
    \ENDIF
\ENDFOR
\RETURN $S$;\\
\end{algorithmic}
\end{algorithm}

\subsubsection{Commitment}
Each user $i$ chooses a bid $b_{i}$ and a limit $l_{i}$ of the
number of assignments to form his sensing profile, 
and then interacts with the AI. According
to the Algorithm \ref{OT}, each user $i$ receives
$\tilde{b}_{i}=\gamma_{x}$ and his limit $\tilde{l}_{i}=\tau_{x}$,
which are the rank-preserving-encrypted values of $\beta_{x}$ and
$\chi_{x}$ respectively, thereby forming his encrypted sensing
profile. Then each user $i$ encrypts the encrypted sensing profile
as $e_{i}=E_{K_{ppub}}(\tilde{b}_{i}|\tilde{l}_{i}|r_{i})$ by using
the platform's Paillier encryption key $K_{ppub}$ and a randomly
chosen values $r_{i}$. User $i$ then makes a commitment
$c_{i}=E_{TPK}(e_{i}|s_{i}|TID)$, where $s_{i}$ is a randomly
generated bit string for the proof of correctness and $TID$ is the
auction identifier ID. Finally, the user signs this commitment, and
sends a bidding request $BR=(sign_{i}(c_{i}|TID))$ to platform, if
used, before time $T$ (see Fig.~\ref{fullframework}, step
\textcircled{\small{1}}). The platform returns a signed receipt
$R_{i}=sign_{p}(c_{i}|TID|T)$ (see Fig.~\ref{fullframework}, step
\textcircled{\small{2}}). At time $T$, the platform posts all the
received true commitments $c_{1},c_{2},\cdots,c_{n}$ on the bulletin
board. 

Note that hiding of the encrypted bids and of the random strings by
applying the secondary encryption prevents anyone from learning any
knowledge of the data prior to time $T$. In particular, neither the
AI nor the platform has any meaningful information.

Furthermore, between time $T$ and $T+1$, for any user who has a
receipt for a bid which is not posted (see Fig.~\ref{fullframework},
step \textcircled{\small{3}}), he can appeal his non-inclusion,
resorting to the AI.

\subsubsection{Decommitment}\label{winningselection}
At time $T+1$, each party, including the platform and all users, can
recover each encrypted sensing profile $e_{i}$ as well as each
random string $r_{i}$ by employing the decryption key TSK posted by
the AI. The platform recovers the pair for computing the auction's
results and random values $r_{1},\cdots,r_{k}$ for the verification
of correctness by applying the platform decryption key. The platform
then computes the set of winners and their corresponding payments
from the platform according to the above auction mechanism with the
given budget constraint. The platform posts the winner's identity
and the encrypted payment information so that any party can verify
the correct results on the bulletin board.

(a)\textbf{Winners Selection:} In this stage, our goal of the
winners' selection is to find the biggest integer $k$ so that $b_{k}
\le B/{\sum\nolimits_{i = 1}^k {f_{i}}}$ holds, thereby obtaining
the set of winners. Firstly, the platform first recovers the bids
$\tilde{b}_{i}$ from the bulletin board and then sorts all users'
encrypted bids from all users and resorts to the AI to fetch the
original value $b_{1}$ of $\tilde{b}_{1}$: $b_{1}=
OPENS^{-1}(\tilde{b}_{1})$. If $b_{i}\leq B/{\sum\nolimits_{j\leq i}
{f_{j}}}$ holds, then users with the rank $1,2,\cdots,i$ are
winners, thus, for the platform, privacy leakage does not exist.
Otherwise, the largest number $k=i-1$. When user $i$ is added to the
set of winners, the platform then computes his assignments
$\tilde{f}_{i}\leftarrow\min\{OPENS^{-1}(\tilde{l}_{i}),
\tilde{\tau}_{i}\}$. The iteration is repeated until our goal is
achieved. The set of winners $\{1,2,\cdots,k\}$ is found. Notable,
when we determine the largest $k$, if $b_{i}\leq
B/{\sum\nolimits_{j\leq i} {f_{j}}}$ does not hold, the $k+1$-th
user's bids and assignments, i.e., its sensing profile, may be
disclosed (see Fig.~\ref{fullframework}, step
\textcircled{\small{4}}). Since in our crowd sensing applications,
we assume that the number of users is much larger than the number
$k$. As such, our scheme satisfies $k$-anonymity. So, neither the
AI, nor the platform, can identify any user's sensing profile with
the probability higher than $1/k$. The detailed description is given
in the Algorithm \ref{winnerpaymentselection}.

(b)\textbf{Payment Decision:} In the payment determination phase,
the platform pays $p_{j}f_{j}$ to user $j$ for $j\leq i$. Similarly,
for each winner $i\in S$, the payment of  per sensing job, i.e.,
$p_{i}$, is given in Algorithm \ref{winnerpaymentselection}. In
particular, our payment scheme is applicable to homogeneous and
heterogeneous sensing job models (see Fig.~\ref{fullframework}, step
\textcircled{\small{5}}).

\begin{figure*}[t]
\setlength{\abovecaptionskip}{0pt}
\setlength{\belowcaptionskip}{10pt} \centering
\centering
\includegraphics[width=0.79\textwidth]{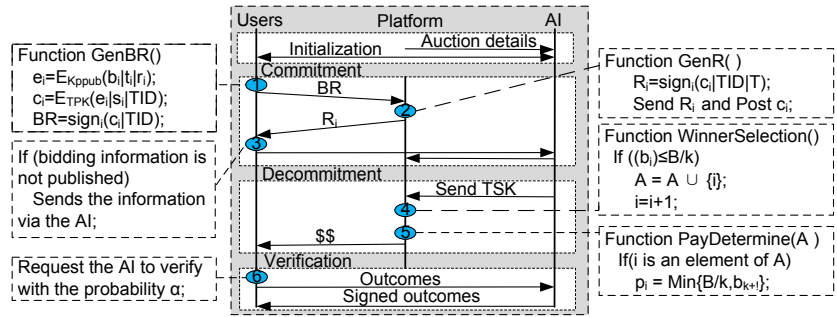}
\caption{Our privacy-preserving verifiable framework for homogeneous
and heterogeneous jobs.} \label{fullframework}
\end{figure*}

\subsubsection{Verification}\label{hhverficaiton}
\begin{figure}[htb]
\setlength{\abovecaptionskip}{0pt}
\setlength{\belowcaptionskip}{10pt} \centering
\centering
\includegraphics[width=0.42\textwidth]{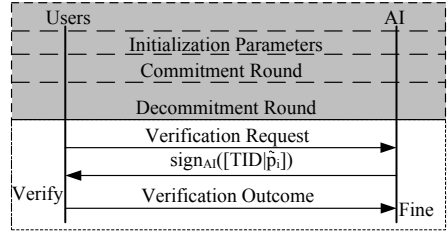}
\caption{Our verifiable phase for homogeneous and heterogeneous
jobs.} \label{verify}
\end{figure}

Since the above incentive mechanism guarantees the truthfulness for
users, we only need to verify the payment correctness of the
platform, that is, any of the users can verify the outcome of the
auction on his own. The detailed descriptions are given as follows.
Firstly, user $i$ requests AI to verify the payment outcome with the
probability $\alpha$. After the AI receives the request, he asks for
the random value $r_{i}$ of each user's $e_{i}$. Then he derives
each user $e_{i}$'s $\tilde{b}_{i}, \tilde{l}_{i}$ by decrypting
$e_{i}$ on the bulletin board with $r_{i}$, thereby obtaining the
payment according to the above auction details and the information
from the bulletin board. He sends the encrypted payment $f_{i}p_{i}$
and his assignments $f_{i}^{'}$ to the user $i$ to verify the
correctness of the outcomes from the
platform,
thereby obtaining the user's feedback to determine whether to fine
the platform (see Fig.~\ref{verify}). 
Analysis in the following section shows that the platform operates
correctly and does not try to cheat.

\subsection{Design Privacy-Preserving Details for Submodular Sensing Jobs}
Different from the above mechanism, for the submodular sensing job
model, we need overcome the challenge of protecting platform's
privacy, i.e., the privacy of the current winners' set, by using the
above-mentioned MPEP method and homomorphic encryption scheme. 
The detailed descriptions are described below.
\subsubsection{Initialization}
The platform invites the AI to participate in the auction and sends
the following information to him: the crowd sensing task identifier
$TID$ of the platform, the description of the mechanism, the
deadline $T$, and the timed-lapse encryption key $TPK$ to be used by
all users in commitments. If the AI accepts them, it sets the
probability of the auditions from the users as $\alpha$ so that
$\alpha\geq p_{max}/(F+p_{max})$, where $p_{max}$ and $F$ are the
maximal payment and fine paid from the platform respectively, and
sends signed $\alpha$ and signed auction details to the platform. If
the platform accepts it, the platform posts them on the bulletin
board. Our mechanism also defines a set of possible marginal
utilities per bid as $\beta = \left\{ {\beta _1 ,\beta _2 , \cdots
,\beta _z } \right\}$, where $\beta _1 < \beta _2 < \cdots < \beta
_z$ holds, and requires that each user $i$'s marginal utility per
bid $\omega_{i}\in \beta$. The AI maps each $\beta _i$ to $\gamma
_i$, while preserving the rank, i.e., satisfying $\gamma _1 < \gamma
_2 < \cdots  < \gamma _n$. Similarly, each user's marginal utility
per bid is transformed by using the order preserving encryption
scheme (OPES) \cite{agrawal2004order} for preserving their ranks. To
this end, we assume that the above AI can bootstrap the crowd
sensing application. Then it constructs three dynamic lists for the
verification of payments' correctness initiated by each user. The
first list $l_{i}^{w}$ for user $i$ is used to put his marginal
utility per bid $\omega_{i}(S)$ for the winner determination phase.
The second list $l_{i}^{p}$ is used to put his marginal utility per
bid $\omega_{i}(\mathcal{T})$ for the payment determination phase.
The last dynamic list $l_{i}^{S}$ is constructed for each winner.
All of the above data are posted on the bulletin board, accompanied
by the platform's signature $sign_{p}$.

\subsubsection{Commitment Round for Winner and Payment Determination}
Each user $i$ initially chooses a bid $b_{i}$ and a subset
$\Gamma_{i}$ of assignments according to his valuation he
preferences. Each user $i$ initially computes his marginal utility
$U_{i}(\emptyset)$, thereby obtaining his marginal utility per bid
$\omega_{i}(\emptyset)$. Then he interacts with the AI by using the
Algorithm \ref{OT}, thereby receiving
$\tilde{\omega}_{i,0}(\emptyset)$, where the subscript $0$ denotes
the cardinality of the current winners' set is equal to $0$, and
$\tilde{\omega}_{i,0}(\emptyset)$ is the rank-preserving-encrypted
value of $\beta_{x}$. Then each user $i$ encrypts it as
$e_{i}=E(\tilde{\omega}_{i,0}|r_{i})$ by using the platform's
Paillier encryption key $K_{pub}$ and a randomly chosen value
$r_{i}$. User $i$ then makes a commit
$c_{i}=E_{TPK}([e_{i}|s_{i}|TID])$, where $s_{i}$ is a randomly
generated bit string for the proof of correctness and $TID$ is the
auction identifier $ID$. Finally, user $i$ signs this commitment
$c_{i}$ and the encrypted value $e_{i}$. Then he adds
$sign_{i}(c_{i})$ to the list $l_{i}^{w}$ on the bulletin board and
sends $sign_{i}(e_{i})$ to platform. Receiving all users' values
$sign_{i}(e_{i})$, the platform decrypts and sorts them, thereby
obtaining the user $i$ with the maximal encrypted marginal utility
per bid. Moreover, the platform enters the following winner
determination phase.

\begin{figure*}
\setlength{\abovecaptionskip}{0pt}
\setlength{\belowcaptionskip}{10pt} \centering
\centering
\includegraphics[width=0.72\textwidth]{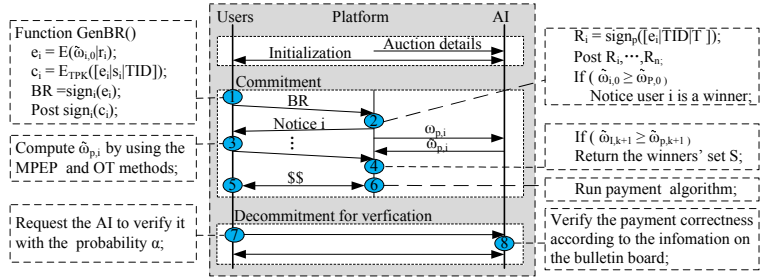}
\caption{Our privacy-preserving verifiable framework for submodular
jobs.} \label{fullframework}
\end{figure*}

(a)\textbf{Winner Determination:} Firstly, the platform applies the
homomorphic encryption scheme to compute the utility $U(S\cup\{i\})$
according to Algorithm \ref{setunion}, thereby obtaining
$\omega_{p,0}=U(S\cup\{i\})/B$. By using the Algorithm \ref{OT}, the
platform interacts with the AI, and receives the encrypted
$\tilde{\omega}_{p,0}$. The platform makes a commit
$c_{p,0}=E_{TPK}([\tilde{\omega}_{p,0}|s_{p}|TID])$, where $s_{p}$
is a randomly generated bit string for the proof of correctness and
$TID$ is the auction identifier $ID$. Signing it,
$sign_{p}(c_{p,0})$, the platform adds it to the list $l_{p}^{w}$ on
the bulletin board. If $\tilde{\omega}_{i,0}(\emptyset)\geq
\tilde{\omega}_{p,0}$, the platform will give user $i$ a notice that
he is a winner. Then the user returns an acknowledgement and his
encrypted $\Gamma_{i}$ and $b_{i}$ by using the platform's public
key. Receiving the acknowledgement, the platform adds user $i$ to
the winners' set $S$ (see the line \ref{fiveline6} of Algorithm
\ref{winnerselection}.) and notifies each user $j\in\mathcal
{U}\setminus S$ to compute his encrypted marginal utility per bid,
i.e., $\tilde{\omega}_{j,1}$, by using the same method as the
computation of $\tilde{\omega}_{i,0}$. These users also add their
signed commitments to their corresponding lists $l_{j}^{w}$ on the
bulletin board. When the platform receives all these
$\tilde{\omega}_{j,1}$, it sorts them, thereby knowing which user
has the maximal encrypted marginal utility per bid. The process is
repeated until the $(k+1)$-th user's
$\tilde{\omega}_{i,k+1}(\emptyset)< \tilde{\omega}_{p,k+1}$.
Finally, we obtain the winners' set that consists of $k$ users.

\begin{algorithm}[htb] 
\renewcommand{\algorithmicrequire}{\textbf{Input:}}
\renewcommand\algorithmicensure {\textbf{Output:} }
\caption{Winner determination for sensing submodular jobs} 
\label{winnerselection} 
\begin{algorithmic}[1] 
\REQUIRE User set $\mathcal {U}$, the budget constraint $B$. \\
\ENSURE The winners' set $S$. \\
\STATE $S\leftarrow \emptyset$; For every $j\in\mathcal {U}$, the
platform recovers $\tilde{\omega}_{j,0}(S)$ by using the decryption
algorithm, and sorts all these values in a decreasing order, thereby
obtaining the user $i$ with the maximal encrypted marginal utility
per bid, i.e.,
$i\leftarrow\arg\max_{j\in\mathcal {U}}\tilde{\omega}_{j,0}(S)$;\\
\STATE The platform obtains $\tilde{\omega}
_{p,0}(S)$ by using Algorithm \ref{OT} and \ref{setunion} and adds a signed commitment to the list $l_{p}^{w}$ on the bulletin board;\\
\WHILE{$\tilde{\omega}_{i,0}(\emptyset)\geq \tilde{\omega}_{p,0}$}
    \STATE The platform notices that user $i$ is a winner;\\
    \STATE Receiving an acknowledgement, the platform adds user $i$ to the winner set $S$, i.e., $S\leftarrow S\cup \{i\}$; \label{fiveline6}\\
    \STATE Notify each user $j\in\mathcal {U}\setminus S$ to compute his encrypted marginal utility per
bid, i.e., $\tilde{\omega}_{j,|S|}$, by using the same method as the
computation of $\tilde{\omega}_{i,0}$; Obtaining all these encrypted
marginal utilities per
bid, the platform finds the user $i$ so that $i\leftarrow\arg\max_{j\in\mathcal {U}\setminus S}(\tilde{\omega}_{j,|S|}(S))$;\\
    \STATE The platform obtains $\tilde{\omega}_{p,|S|}(S)$ by using Algorithm \ref{OT} and \ref{setunion}, and adds a signed commitment to the list $l_{p}^{w}$ on the bulletin board;\\
            \ENDWHILE
            \RETURN $S$;\\
\end{algorithmic}
\end{algorithm}

(b)\textbf{Payment Determination:}\label{Paymentdescription} At this
stage, the encrypted values from the above $OT$ algorithm cannot
support the preserving rank under the multiplication operation. To
address this challenge, we introduce the homomorphic encryption
schemes, which enable multiplication operation of encrypted values
without revealing privacy about the values themselves and the
results of the computation (see equation
(\ref{homomorphicproperty})). Firstly, at time $T$, for each winner
$i\in S$, its payment computation from the platform is given in the
following description. The platform initializes the user set
$\mathcal{U}^{'}$ and set $\mathcal{T}$ by using
$\mathcal{U}^{'}\leftarrow \mathcal{U}\backslash\{i\}$ and
$\mathcal{T}\leftarrow \emptyset$. Differentiating from the above
winner set, we refer to $\mathcal{T}$ as a referenced winner set.
Each user $j\in \mathcal{U}^{'}\backslash\mathcal{T}$ initially
computes his marginal utility $U_{j}(\emptyset)$, thereby obtaining
his encrypted marginal utility per bid
$e_{j,0}=E_{AI}(\omega_{j,0}(\emptyset))$ by using the AI's
homomorphic encryption public key. He makes a commit $c_{j,0}$ by
using the above method. Finally, user $j$ signs this commitment
$c_{j,0}$ and the encrypted value $e_{j,0}$. Then he adds
$sign_{j}(c_{j,0})$ to the list $l_{j,i}^{p}$ on these bulletin
board (meaning that the list is used to put user $j$'s commitment
for the computation of user $i$'s payment) and sends the
$sign_{j}(e_{j,0})$ to the platform. Receiving the values of all
users in $\mathcal{U}^{'}$, the platform sorts them, thereby
obtaining the user $i_{j}$ with the maximal encrypted marginal
utility per bid (i.e., $e_{i_{j},0}$). Then the platform notices
that user $i_{j}$ is a referenced winner and requests user $i$ for
obtaining the $E_{AI}(U_{i(j)})$. After user $i_{j}$ receives the
request, he computes the value $E_{AI}(U_{i(j)})$ by applying the
above MPEP and AI's encryption public key. Signing it, he sends the
signed $E_{AI}(U_{i(j)})$ to the platform. According to the
homomorphic encryption, we have
$E_{AI}(b_{i(j)})=E_{AI}(U_{i(j)}\cdot
b_{i_{j}}/U_{i_{j}})=E_{AI}(U_{i(j)}\cdot1/\omega_{i_{j}})=E_{AI}(U_{i(j)})^{1/e_{i_{j}}}$.
Similarly, we can obtain $E_{AI}(\eta_{i(j)})=E_{AI}(U_{i(j)}\cdot
B/U(\mathcal{T}_{j-1}\cup\{i\}))=E_{AI}(U_{i(j)}\cdot1/\omega_{p,j})=E_{AI}(U_{i(j)})^{1/e_{p,j}}$,
where $e_{p,j}$ means the encrypted marginal utility per bid when
there are $j$ referenced winners. Since user $i$ is a true winner,
the platform knows his bid and sensing preference $\Gamma_{i}$.
Thus, the platform can compute the value $e_{p,j}$. Receiving the
value $E_{AI}(U_{i(j)})$, the platform can obtain $E_{AI}(b_{i(j)})$
and $E_{AI}(\eta_{i(j)})$. Furthermore, the interim payment can be
obtained by using the homomorphic encryption comparison operation.
Subsequently, the platform adds user $i_{j}$ to the referenced
winners' set. The process is repeated until the $(k^{'}+1)$-th
user's $e_{i_{j},k^{'}+1}(\mathcal{T}_{k^{'}})<
e_{p,k^{'}+1}(\mathcal{T})$. Finally, we obtain the payment of
winner $i$. Other winners' payments are computed by adopting the
same method as the winner $i$'s payment. The detailed description is
given in Algorithm \ref{PaymentDecision}.

\begin{algorithm}[htb] 
\renewcommand{\algorithmicrequire}{\textbf{Input:}}
\renewcommand\algorithmicensure {\textbf{Output:} }
\caption{Payment determination for sensing submodular jobs} 
\label{PaymentDecision} 
\begin{algorithmic}[1] 
\REQUIRE User set $\mathcal {U}$, the budget
constraint $B$, the set of winners  $S$. \\
\ENSURE ($\mathcal {U}$, $p$). \\
\FOR{each user $i\in \mathcal {U}$} \STATE $\hat{p}_{i}\leftarrow E_{AI}(0)$;\\
\ENDFOR

\FORALL{user $i\in S$}
\STATE $\mathcal{U}^{'}\leftarrow \mathcal {U}\backslash\{i\}$; the referenced winners' set $\mathcal{T}\leftarrow \emptyset$;\\
\REPEAT
    \STATE Every $j\in\mathcal{U}^{'}$ computes his encrypted
marginal utility per bid
$e_{j,|\mathcal{T}|}=E_{AI}(\omega_{j,|\mathcal{T}|}(\mathcal{T}))$
by using AI's homomorphic encryption public key for sending to the
platform, and adds a signed commitment
$sign_{j}(c_{j,|\mathcal{T}|})$ to the list $l_{j,i}^{p}$ on these
bulletin board; Receiving these encrypted values, the platform sorts
them in a decreasing order, thereby obtaining the user $i_{j}$ with
the maximal encrypted marginal utility per bid, i.e.,
$i_{j}\leftarrow\arg\max_{j\in\mathcal {U}^{'}\backslash\mathcal{T}}e_{j,|\mathcal{T}|}(\mathcal{T})$;\\
    \STATE Notice that user $i_{j}$ is a referenced winner and requests
user $i$ for obtaining the $E_{AI}(U_{i(j)})$; \\

    \STATE According to the description of Section \ref{Paymentdescription}, the platform computes $E_{AI}(b_{i_{j}})$
    and $E_{AI}(\eta_{i(j)})$ by applying equation
(\ref{homomorphicproperty}); Obtain $\hat{p}_{i}\leftarrow \max \{\hat{p}_{i}, \min \{E_{AI}(b_{i_{j}}), E_{AI}(\eta_{i(j)})\}$; \\
   \STATE $\mathcal{T}_{j-1}\leftarrow \mathcal{T}$; $\mathcal{T}\leftarrow \mathcal{T}\cup \{i_{j}\}$; \\
\UNTIL{$e_{i_{j},k^{'}+1}(\mathcal{T}_{k^{'}})<e_{p,k^{'}+1}(\mathcal{T})$
or $\mathcal{T}=\mathcal{U}^{'}$}
\STATE The platform requests the AI for obtaining the payment, i.e., $p_{i}=D_{AI}(\hat{p}_{i})$, where $D_{AI}$ denotes the decryption by using the AI's private key;\\
 \ENDFOR
 \RETURN ($\mathcal {U}$, $p$);\\
\end{algorithmic}
\end{algorithm}
%
%
%
%
%
%
%
%
%
%
%
\subsubsection{Decommitment Round for Verification}\label{winningselection}
Since the mechanism itself is truthful, i.e., each user always
submits his true cost, we only need to demonstrate that any user can
verify the payment correctness of the platform on his own.

\textbf{Verification:} The verification process is similar to the
above description (see Fig.~\ref{verify}). The only difference is
that three dynamic lists in the bulletin board are used to recover
associated values for the payment computation of each user.
Generally speaking, some user initially sends the request of
verification to the AI with the probability $\alpha$. Receiving the
request, the AI runs the algorithm description on the bulletin board
with the help of the values in the three lists until the payment is
obtained. 
For more details of verification, we refer readers to Section
\ref{hhverficaiton} and Fig.~\ref{verify}.

\subsubsection{Privacy-preserving Verifiable
Incentive Mechanism for Sensing Submodular Jobs} In our truthful
privacy-preserving verifiable incentive mechanism for sensing
Submodular jobs, the platform determines a winner $i$'s payment
illustrated in Algorithm \ref{Spva}. At the initial stage, there are
some initial parameters specified by the platform. Then, the
platform performs the winner selection algorithm and the payment
determination algorithm. Once the platform finishes the payment, the
user $i$ will request the AI to verify the platform's payment
correctness with the probability $\alpha$. About the detailed
descriptions of privacy preservation and verification are given in
Algorithm \ref{winnerselection}, Algorithm \ref{PaymentDecision} and
Algorithm \ref{Spva}.

\begin{algorithm}[htb] 
\renewcommand{\algorithmicrequire}{\textbf{Input:}}
\renewcommand\algorithmicensure {\textbf{Output:} }
\caption{PVI-S// Privacy-preserving Verifiable auction mechanism for Crowd sensing application with sensing Submodular jobs} 
\label{Spva} 
\begin{algorithmic}[1] 
\REQUIRE User set $\mathcal {U}$, the budget
constraint $B$. \\
\ENSURE ($\mathcal {U}$, $p$). \\
\STATE Initialize the auction information and encryption tools;\\
\STATE Choose the winners by applying the algorithm ;\\
\STATE Finish the payment for each winner; \\
   \STATE The user requests the AI to verify the payments with the probability $\alpha$;\\
 \RETURN ($\mathcal {U}$, $p$);\\
\end{algorithmic}
\end{algorithm}

\section{Privacy, Verifiability and Revenue Analysis}\label{Analysis}
\subsection{Privacy of Users and Platform}
Our mechanisms' private information include users' privacy and
platform's privacy, i.e., the sensing profile privacy of users and
the current winners' set privacy of the platform. Assume that there
are two kinds of adversaries: adversarial users and adversarial platform or AI. 
The specific analysis is given as follows. 


\begin{lemma0}
The mechanisms PVI-H and PVI-S are privacy-preserving for users.
\end{lemma0}

\begin{proof}
We only need to consider two cases in which the privacy of each user
$i$ may be leaked as follows. The first case is for the adversarial
platform or AI. In the two mechanisms, the platform performs the
winners' selection, and only can know the ($k+1$)-th user's sensing
profile $\mathcal{P}_{k+1}$, but does not know which user it belongs
to. In the stage of verification, similarly, the AI also knows the
($k+1$)-th user's sensing profile $\mathcal{P}_{k+1}$, and does not
know which user it belongs to. The AI and platform only know the
encrypted sensing profile, but have no way to decrypt any of them.
No other party can get even more information than the platform or
AI. On the one hand, user $i$ gets his sensing profile
$\mathcal{P}_{k+1}$ through a $1$-out-of-$z$ OT from the AI, who is
unaware of which sensing profile have been accessed by the user.
User $i$ sends the encrypted sensing profile to the AI, who cannot
decrypt the encrypted sensing profile without knowing the private
key of asymmetric encryption scheme. Even if the AI may know the
($k+1$)-th user's sensing profile later when the platform consults
him, he still cannot infer his user owing to the random number.
Thus, the AI cannot know the user of ($k+1$)-th user. Additionally,
although the platform can obtain the ($k+1$)-th user's sensing
profile later, he can only reversely map the encrypted ($k+1$)-th
user's sensing profile to the original ($k+1$)-th user's sensing
profile with the help of the AI. However, the platform still cannot
derive the user, to which ($k+1$)-th user's sensing profile belongs
out of at least $k$ members according to the Theorem 3.2 in
\cite{singer2010budget} due to a large number of users much larger
than $k$ existing in the crowd sensing applications. Therefore,
neither the AI, nor platform, can know any user's sensing profile
with the probability higher than $1/k$, thereby guaranteeing
$k$-anonymity.

The second case is for an adversarial user. In the two mechanisms,
an adversarial user $j$ does not learn side information during our
mechanisms no matter he is a winner or not. All he learns from the
two mechanisms are included in the valid auction's $Output$, i.e.,
for an adversarial user $j$'s advantage $adv_{\mathcal{P}_{i}}$ are
all equal to $0$ for all $i\neq j$.

Putting them together, the lemma holds.
\end{proof}

Besides, in the following lemma, we also analyze the privacy
preservation performance of the platform.
\begin{lemma0}
For the current winners' set $S$ and referenced winners' set
$\mathcal{T}$ of the platform (the privacy of the platform), an
adversarial user $j$'s advantage, i.e., $adv_{S}$ and
$adv_{\mathcal{T}}$, are equal to $0$. In other words, the
mechanisms PVI-H and PVI-S are privacy-preserving.
\end{lemma0}
\begin{proof}
For the current winners' set $S$ and referenced winners' set
$\mathcal{T}$, only platform and AI learn the two sets and each user
learns nothing. Since the AI is semi-honest, and only check the
platform randomly, adversarial users gain no useful information on
the two sets from the communication strings. Thus, the priori
probability is same as the posterior probability, i.e.,
$Adv_{S}=P_{r}[S|\mathcal{C},Output]-P_{r}[S|Output]=0$ and
$Adv_{\mathcal{T}}=P_{r}[\mathcal{T}|\mathcal{C},Output]-P_{r}[\mathcal{T}|Output]=0$.
Thus, the mechanisms PVI-H and PVI-S are privacy-preserving for the
platform. Thus, the lemma holds.
\end{proof}

Putting these lemmas together, we have the following theorem.

\begin{theorem0}
The mechanisms PVI-H and PVI-S are privacy-preserving.
\end{theorem0}

\subsection{Verifiable Correctness of Payments}

\begin{lemma0}\label{sampleofeffort}
The users in the mechanisms PVI-H and PVI-S is truthful.
\end{lemma0}
\begin{proof}
For the mechanism PVI-H, we can easily extend the outcome of the
homogenous jobs presented by Singer et al. \cite{singer2010budget}
the proof outcome to the heterogeneous jobs. For the mechanism PVI-S
according to \cite{yang2012crowdsourcing}, since Algorithm PVI-S is
designed based on the MSensing mechanism of
\cite{yang2012crowdsourcing}, they have demonstrated the
truthfulness of the mechanism, our mechanism PVI-S is also truthful
for users in crowd sensing applications. Thus, the lemma holds.
\end{proof}

Generally speaking, the verifiability issue includes the
Verifiability of users' sensing profile and platform's payment. From
the above lemma \ref{sampleofeffort}, we know that users' bid is
truthful. Besides, each user's subset of assignments is fixed in our
mechanisms. Thus, each user's sensing profile is truthful.
Therefore, we only need to guarantee the verifiable correctness of
payments from the platform. Furthermore, we have the following
lemma.

\begin{lemma0}\label{guessbids}
The two proposed mechanisms, i.e. PVI-H and PVI-S, are correct for a
rational platform.
\end{lemma0}
\begin{proof}
Correctness of both PVI-H and PVI-S, follows from the assumption
that the assumption that the platform is rational and the fine that
he pays when checked cheating is high enough. If his expected
utility when complying with both PVI-H and PVI-S is higher than his
expected utility from his deviation he will abide by the algorithm,
as such the proposed algorithms i.e. PVI-H and PVI-S, will be
correct. We will show the probability $\alpha$ that the platform's
incorrect payment will not be checked by the user with the help of
the AI, set by the two algorithms i.e. PVI-H and PVI-S, ensures that
the platform's expected utility is non-positive \cite{catanesecure}.
The detailed derivation is given as follows. $\alpha\geq
p_{max}/(f+p_{max})\Rightarrow (1-\alpha)p_{max}-\alpha f\leq0$.
Considering the platform's expected utility, i.e., $(1-\alpha)V_{+}
+\alpha V_{-}$, where $ V_{+}$ denotes the platform's utility when
it gives incorrect payment but is not checked by the users, and $
V_{-}$ denotes the platform's utility when it gives incorrect
payment but is checked by the users. Again, $p_{max}\geq V_{+}$ and
$-f= V_{-}$, according to the outcome of the above derivation,
further, we have $(1-\alpha)V_{+} +\alpha V_{-}\leq
(1-\alpha)p_{max}-\alpha f\leq0$. Thus, if the platform does not
comply with the algorithm PVI-H, its expected utility is
non-positive. As such, for a rational platform, it is willing to
abide by the rules of both PVI-H and PVI-S, and gives a correct
payment for every user. Finally, the lemma holds.
\end{proof}

Putting these lemmas together, we have the following theorem.
\begin{theorem0}
The mechanisms PVI-H and PVI-S are verifiable correctness of
payments.
\end{theorem0}
\subsection{Revenue of Platform}
\begin{lemma0}\label{revenue}
The mechanisms in Section \ref{MecahnismDesgn} are
$O(1)$-competitive in maximizing the revenue of the platform.
\end{lemma0}
\begin{proof}
To quantify the revenue of the platform running the mechanisms in
Section \ref{MecahnismDesgn}, we compare their revenue with the
optimal revenue: the revenue obtainable in the offline scenario
where the platform has full knowledge about users' sensing profiles.
A mechanism is $O(1)$-competitive if the ratio between the
mechanism's revenue and the optimal revenue is a constant factor
approximation. According to the Theorem $3.4$ in
\cite{singer2010budget} and Theorem $4.5$ in
\cite{singer2010budget}, we know that the mechanisms in Section
\ref{MecahnismDesgn} are budget feasible constant-approximation
mechanisms, and no budget feasible mechanism can do better than
mechanisms of Section \ref{MecahnismDesgn} in maximizing the
homogeneous, heterogeneous sensing revenue and submodular sensing
revenue of the platform. Thus, the lemma holds.
\end{proof}

Furthermore, different from the mechanisms in Section
\ref{MecahnismDesgn}, mechanisms PVI-H and PVI-S mainly apply the
order preserving encryptions and the OT operations. However, these
encryptions and operations in mechanisms PVI-H and PVI-S do not
change the allocation and payment rules of the mechanisms in Section
\ref{MecahnismDesgn}. Thus, mechanisms PVI-H and PVI-S keep the same
revenue as the mechanisms in Section \ref{MecahnismDesgn}, thereby
obtaining the following theorem.
\begin{theorem0}
The mechanisms PVI-H and PVI-S achieve the same revenue as the
generic one without privacy preservation.
\end{theorem0}

\section{Performance Evaluation}~\label{Experiment}
In this section, we analyze the communication and computation
overhead to show our construction is both scalable and efficient,
thereby applying to mobile devices for crowd sensing applications.
Most of the complexities are linear to the number of users or the
number of assignments, which allows huge number of users or the
number of assignments. Meanwhile, the extra data transmission and
the run time introduced by our mechanisms are almost negligible.
\subsection{Simulation Setup}
To better evaluate the computation overhead, we implemented the PVI
mechanism in Ubuntu 12.04 using the GMP library based on $C$ in a
computer with Intel(R) Core(TM)i5-3470 CPU 3.20GHz. 
The order $p$ of the integer group $\mathbb{Z}_{p}$ is selected as a
$1024$-bit prime number, and users can get $128$ bits of
order-preserving encrypted value through oblivious transfer with the
AI. Every operation is run $100$ times to measure the average run
time.
\subsection{Performance Evaluation for The PVI-H Mechanism}

\subsubsection{Bulletin Board Storage Complexity} We require the
bulletin board to store the auction details and dynamic lists used
to store the parameters or values accessed by the platform and AI.
In each list, there are only few elements. Therefore, the storage
complexity is $\theta(n)$, where $n$ is the number of users.
\subsubsection{Communication Overhead} The communication overhead
based on the data transmission is illustrated in Table
\ref{tab:CommunicationOverhead1}, where $l_{bit}$ is the bit length
of the $p$ (i.e. the order of the integer group $\mathbb{Z}_{p}$).

Note that the computation of accumulated assignments in the winner
determination phase is executed until the platform finds the largest
$k$ so that $b_{i}\leq B/{\sum\nolimits_{j\leq i} {t_{j}}}$ holds.
Thus, the average communication rounds for the platform should be
much less than $O(mn)$, which means that the real communication
overhead will be much less than the worst case $O(mnl_{bit})$. Since
the verification from the AI does not need the communication for the
computation of accumulated assignments, it only requires information
from the existing bulletin board. Thus their communication overhead
is negligible. Fig.~\ref{communication} shows that the overall
communication overhead induced by Algorithm PVI-H. Obviously, the
communication overhead is mainly from the OT.

\begin{table}[h]
\centering \tbl{Communication Overhead of
PVI-H\label{tab:CommunicationOverhead1}}{

\begin{tabular}{|c|c|c|c|c|}
\cline{1-5}  \multicolumn{5}{|c|}{Winner Selection} \\
\hline & Send & Receive & $\omega _{i}$ computation & User sorting\\
\hline Users &$O(n)$  &$O(n)$ & $O(n)$ &0\\
\hline Platform & $O(nm)$ & $O(n)$ &0  &$O(n^{2})$\\
\hline AI & $O(n)$ & $O(n)$ & 0 &$O(n^{2})$\\
\cline{1-5}  \multicolumn{5}{|c|}{Payment Determination} \\
\hline Each winner & 0 & $O(1)$ & $O(n)$ & 0\\
\hline Platform & $O(m^{2})$ & $O(m)$ & 0 &$O(m^{2})$\\
\hline AI & $O(m)$ & $O(m)$ &0  &$O(m^{2})$\\
\cline{1-5}  \multicolumn{5}{|c|}{Verification} \\
\hline Each winner  & $O(1)$ & $O(1)$ & 0 & 0\\
\hline Platform & $O(n)$ & $O(n)$ & 0 &$O(n^{2})$\\
\hline AI & $O(n)$ & $O(n)$ &0  &$O(n^{2})$\\
\hline
\end{tabular}}
\end{table}

\subsubsection{Computation Overhead} In theory, the overhead of the
computation is summarized in Table \ref{tab:ComputationOverhead2}.

In the PVI-H mechanism, the auction is composed of the winner
selection phase and the payment determination phase. The winner
selection phase mainly includes the OT, the sorting, users' blind
signature generation and the computation of accumulated assignments;
the payment determination phase mainly includes the payment
calculation of the platform. For each user's verification, since
data applied to verify the payment are stored in the bulletin board,
the computation overhead of the verification is negligible when
compared with the above parts. Thus, we do not account for it. Next,
we analyze their run time in turn.

\begin{table}[h]
\centering \tbl{Computation Overhead of
PVI-H\label{tab:ComputationOverhead2}}{
\begin{tabular}{|c|c|c|c|}
\hline & Winner selection & Payment determination & Verification\\
\hline Users &$O(1)$  &$O(1)$ & $O(1)$ \\
\hline Platform & $O(nm^{2})$ & $O(nm^{3})$ &0 \\
\hline AI & $O(1)$ & $O(1)$ & $O(nm^{3})$ \\
\hline
\end{tabular}}
\vspace{-5pt}
\end{table}

(a)\textbf{Blind Signature Generation:} In the PVI-H mechanism, the
signer's run time for the user's one is 19 ms and blindly generating
one pair of the Nyberg-Rueppel signature is 11 $\mu s$(microseconds)
on average.


(b)\textbf{Calculation of Assignments and Payment:}

Since a single calculation needs $0.4 \mu s$ on average, and the
overall computation overhead is very small, the run time of the
calculation of accumulated assignments and payment for various
number of assignments and payment is almost negligible.
%

%

\subsection{Performance Evaluation for The PVI-S Mechanism}
\subsubsection{Bulletin Board Storage Complexity} We require the
bulletin board to store the auction details and three dynamic lists
of each user used to store the values accessed by the AI. In each
list, there are only few elements. Therefore, the storage complexity
is $\theta(n)$, where $n$ is the number of users.
\subsubsection{Communication Overhead}

The communication overhead in terms of transmitted bits is
summarized in Table \ref{tab:CommunicationOverhead3}. Note that the
marginal-utility-per-bid computation in the winner selection and
payment determination is executed until the platform and AI finish
the winner selection and the payment determination. Because there
are $m$ different assignments, and each winner should contribute at
least one new assignment to be chosen, the number of winners in the
payment determination phase is at most $m$. Thus, the average
communication rounds for the platform should be much less than
$O(m^{2})$, which means the practical communication overhead will be
much less than the worst case $O(m^{2})$. Besides, the MPEP's
introduction for the marginal-utility-per-bid computation, makes the
communication overhead of each user different with the PVI-H
mechanism (see Table \ref{tab:CommunicationOverhead3}).

\begin{table}[h]
\centering \tbl{Communication Overhead of
PVI-S\label{tab:CommunicationOverhead3}}{

\begin{tabular}{|c|c|c|c|c|}
\cline{1-5}  \multicolumn{5}{|c|}{Winner Selection} \\
\hline & Send & Receive & $\omega _{i}$ computation & User sorting\\
\hline Users &$O(n^{2})$  &$O(n^{2})$ & $O(n)$ &0\\
\hline Platform & $O(nm)$ & $O(n)$ &0  &$O(n^{2})$\\
\hline AI & $O(n)$ & $O(n)$ & 0 &$O(n^{2})$\\
\cline{1-5}  \multicolumn{5}{|c|}{Payment Determination} \\
\hline Each winner & 0 & $O(1)$ & $O(n)$ & 0\\
\hline Platform & $O(m^{2})$ & $O(m)$ & 0 &$O(m^{2})$\\
\hline AI & $O(m)$ & $O(m)$ &0  &$O(m^{2})$\\
\cline{1-5}  \multicolumn{5}{|c|}{Veification} \\
\hline Each winner  & $O(1)$ & $O(1)$ & 0 & 0\\
\hline Platform & $O(n)$ & $O(n)$ & 0 &$O(n^{2})$\\
\hline AI & $O(n)$ & $O(n)$ &0  &$O(n^{2})$\\
\hline
\end{tabular}}
\vspace{-5pt}
\end{table}

\subsubsection{Computation Overhead} In theory, the overhead of the
computation is summarized in Table \ref{tab:ComputationOverhead4}.

\begin{table}[h]
\centering \tbl{Computation Overhead of
PVI-S\label{tab:ComputationOverhead4}}{
\begin{tabular}{|c|c|c|c|}
\hline & Winner selection & Payment determination & Verification\\
\hline Users &$O(1)$  &$O(1)$ & $O(1)$ \\
\hline Platform & $O(nm^{2})$ & $O(nm^{3})$ &0 \\
\hline AI & $O(1)$ & $O(1)$ & $O(nm^{3})$ \\
\hline
\end{tabular}}
\vspace{-5pt}
\end{table}

In general, the PVI-S mechanism consists of the winner selection
phase, the payment determination phase, and the verification phase.
The winner selection phase includes users' blind signature, the
sorting of the platform and the computation of marginal utility per
bid. The payment determination phase includes the sorting of the
platform and the computation of marginal utility per bid. In the
verification phase, since data applied to verify the payment are
stored in the bulletin board, the computation overhead of the
verification is negligible when compared with the above parts. Thus,
we do not account for it. Now, we analyze their run times
respectively.

\begin{figure*}
\center \hspace{-0.08in} \subfigure[]{ \label{communication}
\includegraphics[width=2.10in]{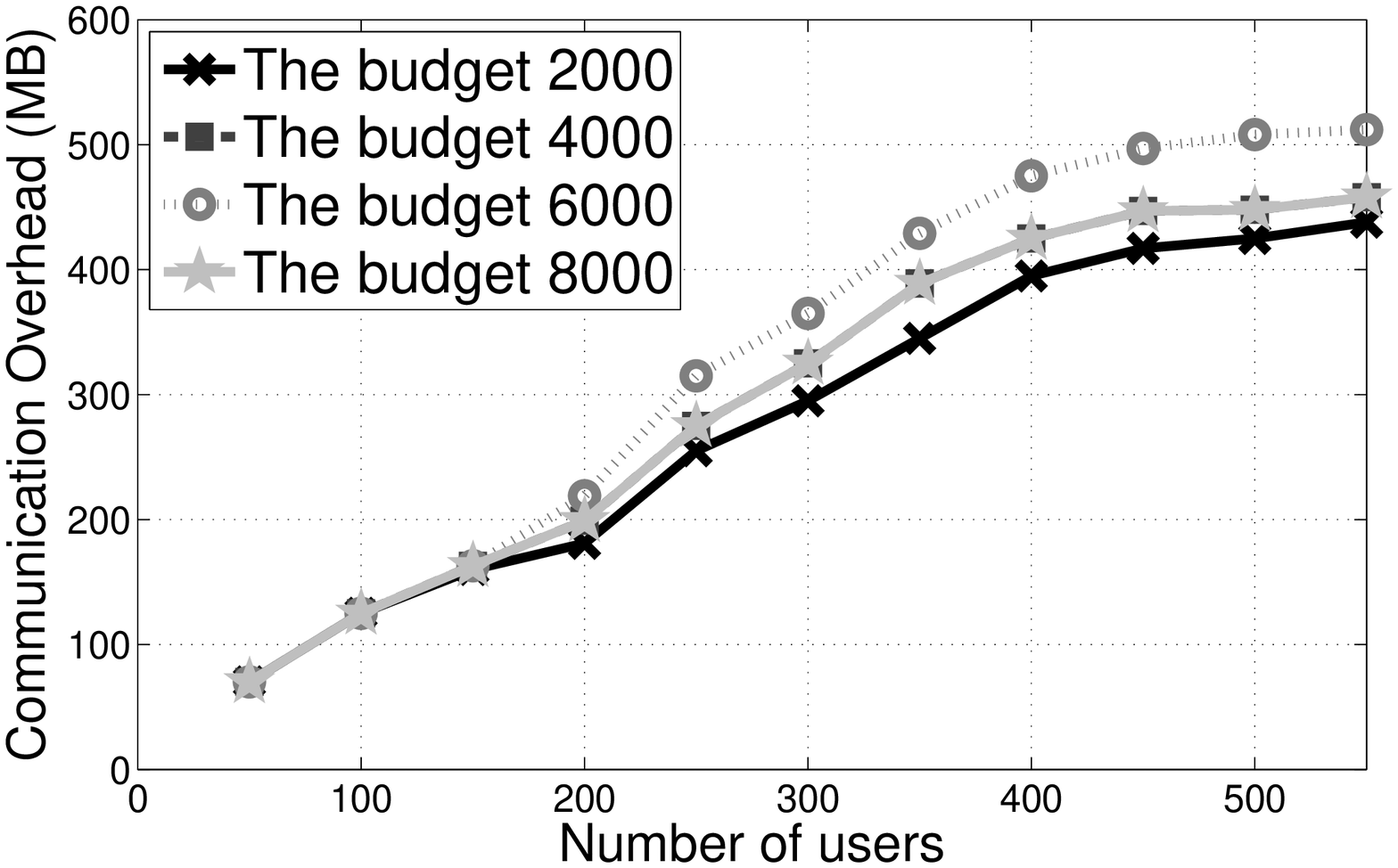}}
\hspace{0.3in} \subfigure[]{ \label{OTsortingsub}
\includegraphics[width=2.10in]{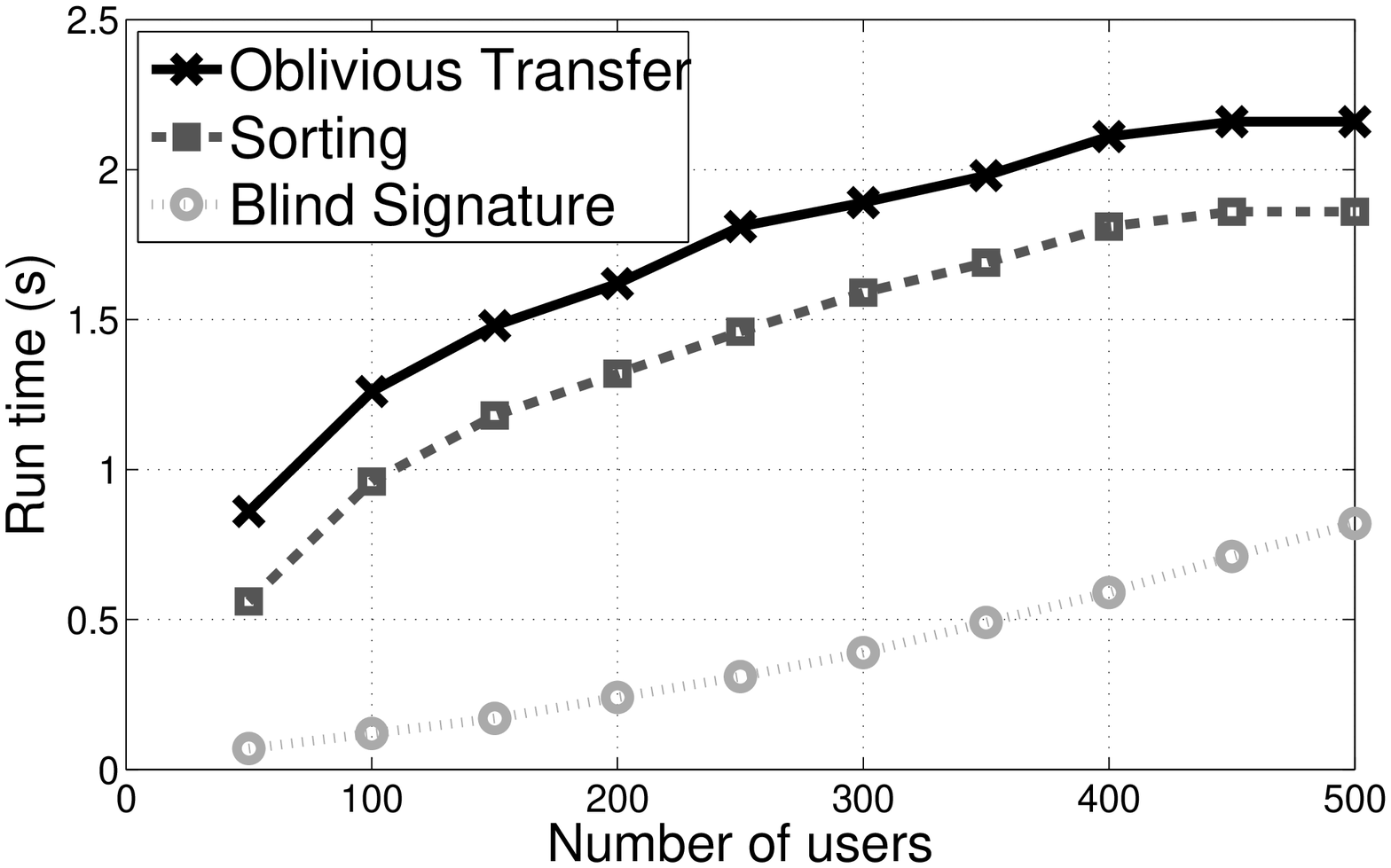}}
\hspace{0.1in} \subfigure[]{ \label{computationsub}
\includegraphics[width=2.10in]{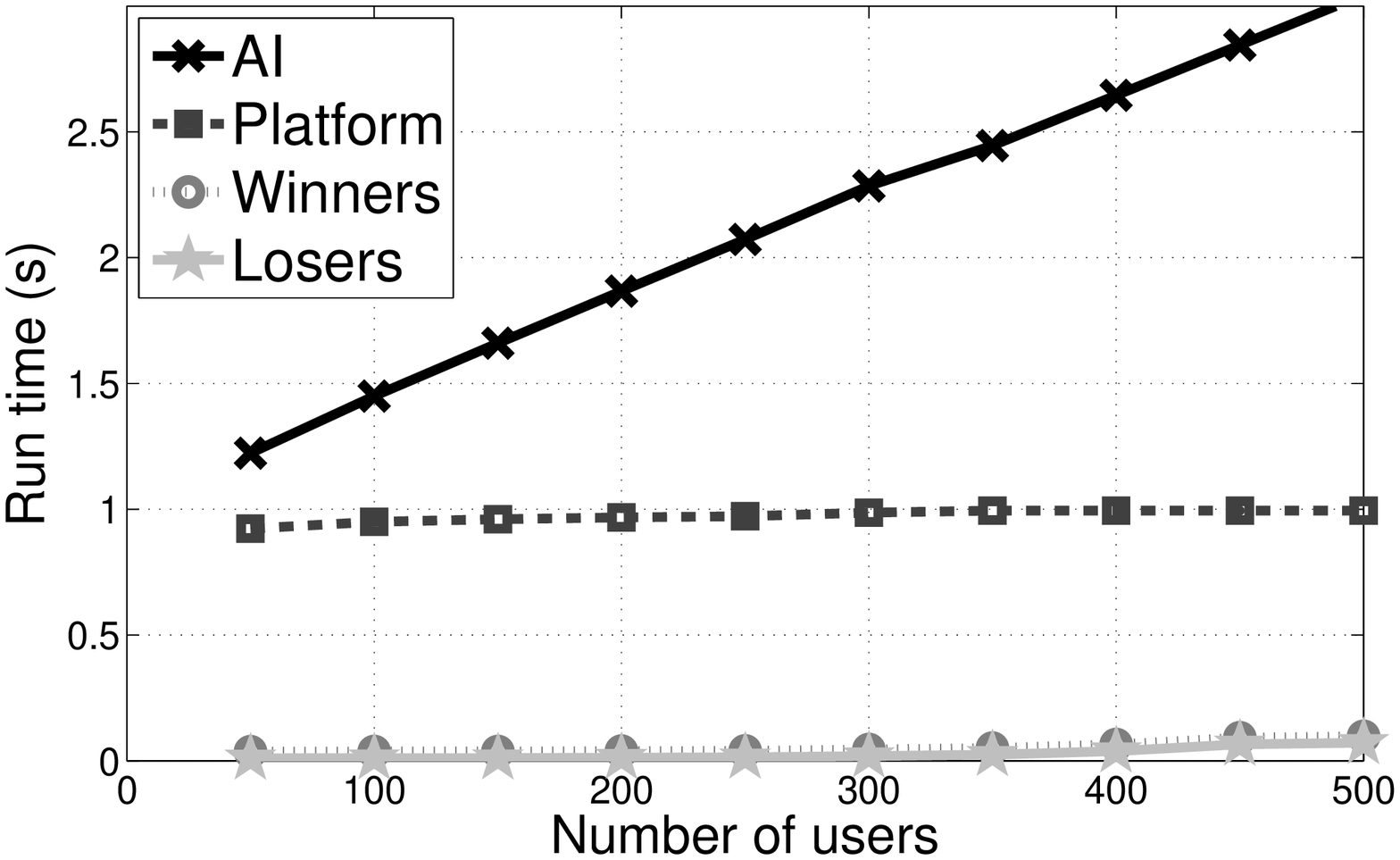}}
\hspace{0.30in} \subfigure[]{ \label{winnersub}
\includegraphics[width=2.1in]{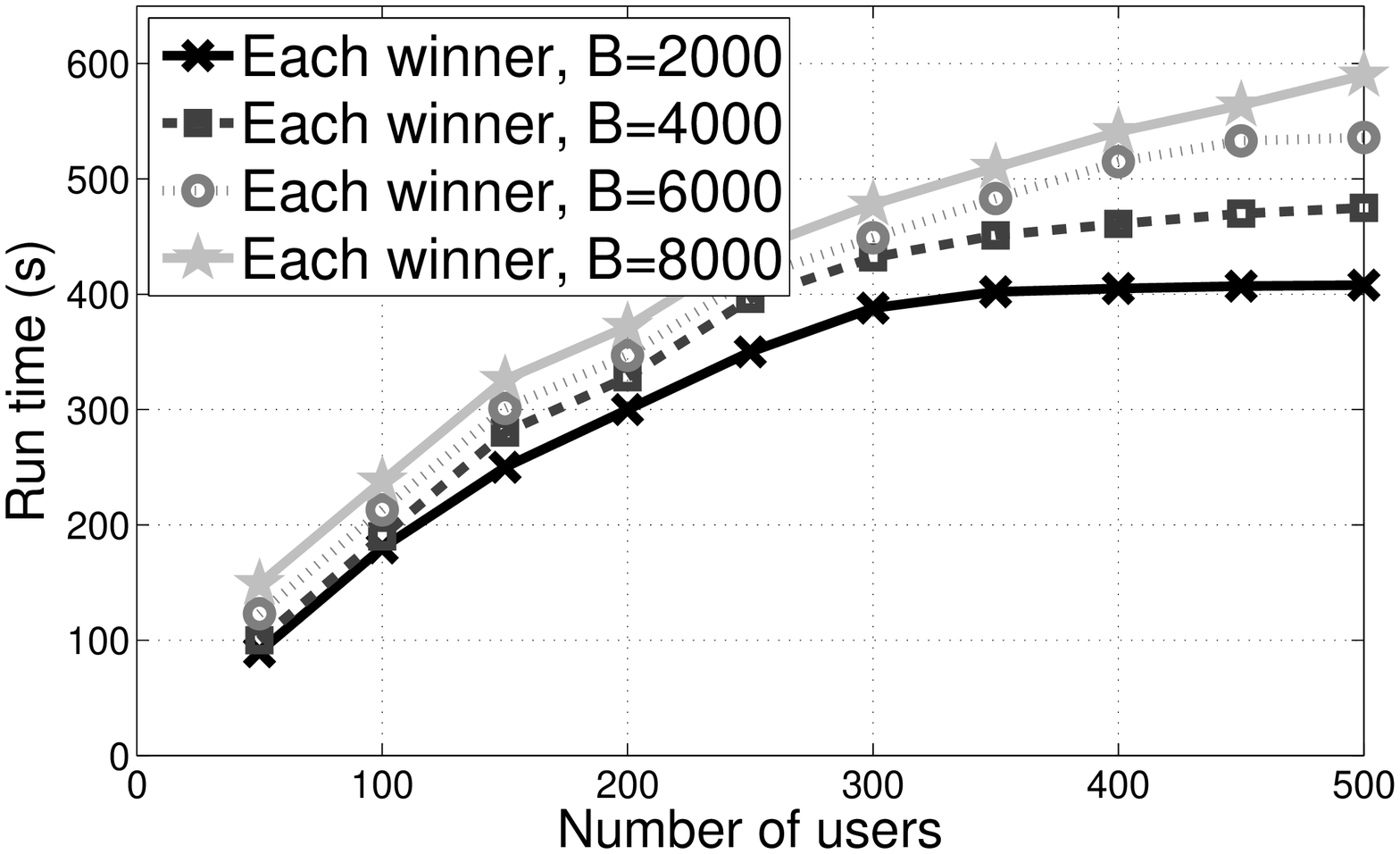}}
\caption{(a)Communication overhead of PVI-H with different budgets;
(b) Run time of the sorting, the OT and the blind Signature of PVI-S
with the number of users when the value of the budget is 2000;
(c)Run time of the AI, the platform, losers and winners of PVI-S
with the number of users when the value of the budget is 2000;
(d)Effect of budget constraint of PVI-S on
computation overhead.} \label{submodelexperiment}
\end{figure*}

(a)\textbf{Sorting, OT and Blind Signature:} The PVI-S mechanism's
run time for one pair of the Nyberg-Rueppel signature including the
AI, platform and users is 28 milliseconds on average. Further, we
also evaluated the run time of the OT as well as the final sorting
based on the encrypted values. We observed that the computation
overhead of the signature is negligible when compared with the one
of the OT and sorting. Users in the PVI-S have much less run time
since they only generate the communication strings (ciphertexts)
(see Fig.~\ref{OTsortingsub}).

(b)\textbf{Computation of AI, Platform, Winners and Losers:} We
compared the computation overhead of the AI, the platform, winners
and losers in Fig.~\ref{computationsub} when the budget value is
$2000$. We observed that the computation overhead increases with the
budget constraint and at last they were kept in a stable constant
value respectively. It is because that at this moment the PVI-S
mechanism reached saturation point.

(c)\textbf{Effect of Budget Constraint on Computation Overhead:} To
assess the effect of different budget constraints on computation
overhead of each winner $i$, we calculated the average computation
overhead of each winner for different budget values respectively. We
observed that the overall computation overhead increased with the
number of winners at last reached a stable value (see
Fig.~\ref{winnersub}). The computation overhead of each user is very
small, therefore the overhead induced by the PVI-S mechanism also
can be applied to wireless mobile devices for crowd sensing
applications.

\section{Concluding Remarks}~\label{Conclude}
In this paper, we design two privacy-preserving verifiable incentive
mechanisms for crowd sensing applications. We not only address the
privacy preservation of users and the platform by applying the OPES
and OT, but also provide a verification scheme for the payment
correctness from the platform by using the signature technology and
the bulletin board. We preserve the rank of the encrypted values by
using the OPES scheme. Furthermore, we prevent bid repudiation by
employing a TLC service. No party, including the platform, receives
any information about bids before the mechanism closes, and no user
is able to change or repudiate any sensing profile. Finally, we
design and analyze the two mechanisms. Results from theory analysis
and experiments indicate that our privacy-preserving verifiable
incentive mechanisms achieve the same results as the generic one
without privacy preservation and apply for mobile devices in crowd
sensing applications. As such, they can apply generally or be
extended to other truthful incentive mechanisms for real crowd
sensing environments.
%

\bibliographystyle{ACM-Reference-Format-Journals}
\bibliography{ACMtran}

\received{December 2013}{March 2014}{June 2014}



%
%
%
%

\end{document}